\documentclass[aps,prb,twocolumn,a4,showpacs,superscriptaddress]{revtex4-1} 
\usepackage{amsmath,amssymb,amsfonts,upgreek}

\usepackage[english]{babel}
\usepackage[utf8x]{inputenc}
\usepackage[T1]{fontenc}
\usepackage{bm}
\usepackage{nicefrac}

\usepackage{rotating}
\usepackage[colorinlistoftodos]{todonotes}
\usepackage[colorlinks=true, allcolors=blue]{hyperref}
\usepackage{relsize}

\usepackage{amsmath,amssymb,amsfonts,upgreek}
\usepackage{gensymb}
\usepackage{color}
\usepackage{graphicx}
\usepackage{setspace}
\usepackage{multirow}
\usepackage{hhline}
\usepackage{bm}
\usepackage{comment}
\usepackage{natbib}

\begin{document}

\title{Relationship between crystal structure and multiferroic orders in orthorhombic perovskite manganites}

\author{Natalya S.\ Fedorova}
\email{natalya.fedorova@mat.ethz.ch}
\affiliation{Materials Theory, ETH Z\"{u}rich, Wolfgang-Pauli-Strasse 27, CH-8093 Z\"{u}rich, Switzerland}
\author{Yoav William Windsor}
\affiliation{Swiss Light Source, Paul Scherrer Institut, CH-5232 Villigen PSI, Switzerland}

\author{Christoph Findler}
\affiliation{Materials Theory, ETH Z\"{u}rich, Wolfgang-Pauli-Strasse 27, CH-8093 Z\"{u}rich, Switzerland}
\author{Mahesh Ramakrishnan}
\affiliation{Swiss Light Source, Paul Scherrer Institut, CH-5232 Villigen PSI, Switzerland}

\author{Amad\'{e} Bortis}
\affiliation{Laboratory for Multifunctional Ferroic Materials, ETH Z\"{u}rich, Vladimir-Prelog-Weg 4, CH-8093 Z\"{u}rich, Switzerland}

\author{Laurenz Rettig}
\affiliation{Swiss Light Source, Paul Scherrer Institut, CH-5232 Villigen PSI, Switzerland}
\author{Kenta Shimamoto}
\affiliation{Laboratory for Multiscale Materials Experiments, Paul Scherrer Institut, CH-5232 Villigen PSI, Switzerland}
\author{Elisabeth M. Bothschafter}
\affiliation{Swiss Light Source, Paul Scherrer Institut, CH-5232 Villigen PSI, Switzerland}

\author{Michael Porer}
\affiliation{Swiss Light Source, Paul Scherrer Institut, CH-5232 Villigen PSI, Switzerland}
\author{Vincent Esposito}
\affiliation{Swiss Light Source, Paul Scherrer Institut, CH-5232 Villigen PSI, Switzerland}
\author{Yi Hu}
\affiliation{Laboratory for Multiscale Materials Experiments, Paul Scherrer Institut, CH-5232 Villigen PSI, Switzerland}
\author{Aurora Alberca}
\affiliation{Swiss Light Source, Paul Scherrer Institut, CH-5232 Villigen PSI, Switzerland}
\author{Thomas Lippert}
\affiliation{Laboratory for Multiscale Materials Experiments, Paul Scherrer Institut, CH-5232 Villigen PSI, Switzerland}
\affiliation{Laboratory of Inorganic Chemistry, Department of Chemistry and Applied Biosciences, ETH Zürich, CH-8093, Z\"{u}rich, Switzerland}
\author{Christof W. Schneider}
\affiliation{Laboratory for Multiscale Materials Experiments, Paul Scherrer Institut, CH-5232 Villigen PSI, Switzerland}
\author{Urs Staub}
\affiliation{Swiss Light Source, Paul Scherrer Institut, CH-5232 Villigen PSI, Switzerland}
\author{Nicola A.\ Spaldin}
\email{nicola.spaldin@mat.ethz.ch}
\affiliation{Materials Theory, ETH Z\"{u}rich, Wolfgang-Pauli-Strasse 27, CH-8093 Z\"{u}rich, Switzerland}

\begin{abstract}
We use resonant and non-resonant X-ray diffraction measurements in combination with first-principles electronic structure calculations and Monte Carlo simulations to study the relationship between crystal structure and multiferroic orders in the orthorhombic perovskite manganites, o-\textit{R}MnO$_3$ (\textit{R} is a rare-earth cation or Y). In particular, we focus on how the internal lattice parameters (Mn-O bond lengths and Mn-O-Mn bond angles) evolve under chemical pressure and epitaxial strain, and the effect of these structural variations on the microscopic exchange interactions and long-range magnetic order. We show that chemical pressure and epitaxial strain are accommodated differently by the crystal lattice of o-$R$MnO$_3$, which is key for understanding the difference in magnetic properties between bulk samples and strained films. Finally, we discuss the effects of these differences in the magnetism on the electric polarization in o-\textit{R}MnO$_3$.
\end{abstract}

\maketitle

\section{Introduction}
The last two decades have seen major activity in the study of magnetoelectric multiferroics, an exciting class of materials that exhibit ferroelectric polarization alongside magnetic order.  Interest in these materials largely stems from the possibility of controlling one order using the stimulus that usually controls the other, offering great potential for development of novel multifunctional devices \cite{spaldin2005multiferroics,spaldin2010MF_past_present}. Among single phase multiferroics, interesting candidates for future technological applications are those in which ferroelectricity is induced by inversion-symmetry-breaking magnetic order (multiferroics of type II) \cite{cross1978magnetoferroelectricity}, since their ferroelectric (magnetic) properties can be easily tuned by applied magnetic (electric) field \cite{cheong2007multiferroics,kimura2003magnetic}. Type II multiferroics are usually frustrated magnets in which competing exchange interactions give rise to several magnetic phases with similar energies. As a result, transitions between them can be driven by control parameters such as chemical or hydrostatic pressure, epitaxial strain, or even by ultrashort light pulses \cite{johnson2012CuO,Bothschafter2017TbMnO3}, offering multiple routes to manipulating and controlling their properties \cite{fiebig2016multiferroics}. 

The orthorhombic \textit{R}MnO$_3$ (o-\textit{R}MnO$_3$), in which \textit{R} is a rare-earth cation or Y, are prototypical representatives of type II multiferroics. It was discovered in 2003 \cite{kimura2003magnetic} that in bulk o-TbMnO$_3$ the establishment of an incommensurate spiral magnetic order \cite{kenzelmann2005tbmno3} gives rise to a spontaneous electric polarization whose direction and magnitude can be manipulated by an external magnetic field. This effect, however, occurred at quite low temperatures and the measured values of the electric polarization were relatively small compared to those of conventional ferroelectrics. 
Nevertheless, this discovery stimulated experimental and theoretical studies aiming to understand and improve the multiferroic properties of systems with frustrated magnetic orders. 
In particular, it was theoretically predicted that E-type antiferromagnetic order (E-AFM), which was observed in early neutron diffraction measurements in o-HoMnO$_3$ \cite{munoz2001homno3} and expected to be a magnetic ground state in other o-\textit{R}MnO$_3$ with small \textit{R},  may induce an electric polarization at least one order of magnitude higher than that of spiral-order systems  \cite{sergienko2006ferroelectricity,picozzi2007dual}. 
The experimental verification of this prediction, however, gave contradictory results. On one hand, the predicted polarization values have not yet been measured experimentally for bulk o-\textit{R}MnO$_3$ \cite{feng2010homno3,lorenz2007homno3_ymno3,chai2012rmno3_polar}. Moreover, magnetic orders different from E-AFM were reported for $R$=Ho, Er, Y \cite{lee2011mechanism,ye2007incommensurate} and there is still no agreement about the type of these orders, the mechanisms of their establishment, or the directions and magnitudes of the electric polarizations which they induce. 
On the other hand, increased polarization values were observed in structurally modified o-\textit{R}MnO$_3$ samples.
Indeed, it has been shown that the spiral order in o-TbMnO$_3$ can evolve to E-AFM under isotropic pressure \cite{makarova2011pressure} and this evolution significantly enhances the electric polarization in this system \cite{aoyama2014tbmno3}. Variations of the magnetic modulation vector and enhancement of electric polarization were also observed in epitaxially strained films of o-$R$MnO$_3$ \cite{wadati2012origin,shimamoto2016multiferroic,shimamoto2017phase_diagram}. 
The microscopic origin of such an evolution of the magnetic order in strained samples as well as the difference in magnitudes of the electric polarization between bulk and strained samples of o-$R$MnO$_3$, however, are still not understood. 

\begin{figure}
\centering
\includegraphics[width=0.5\textwidth]{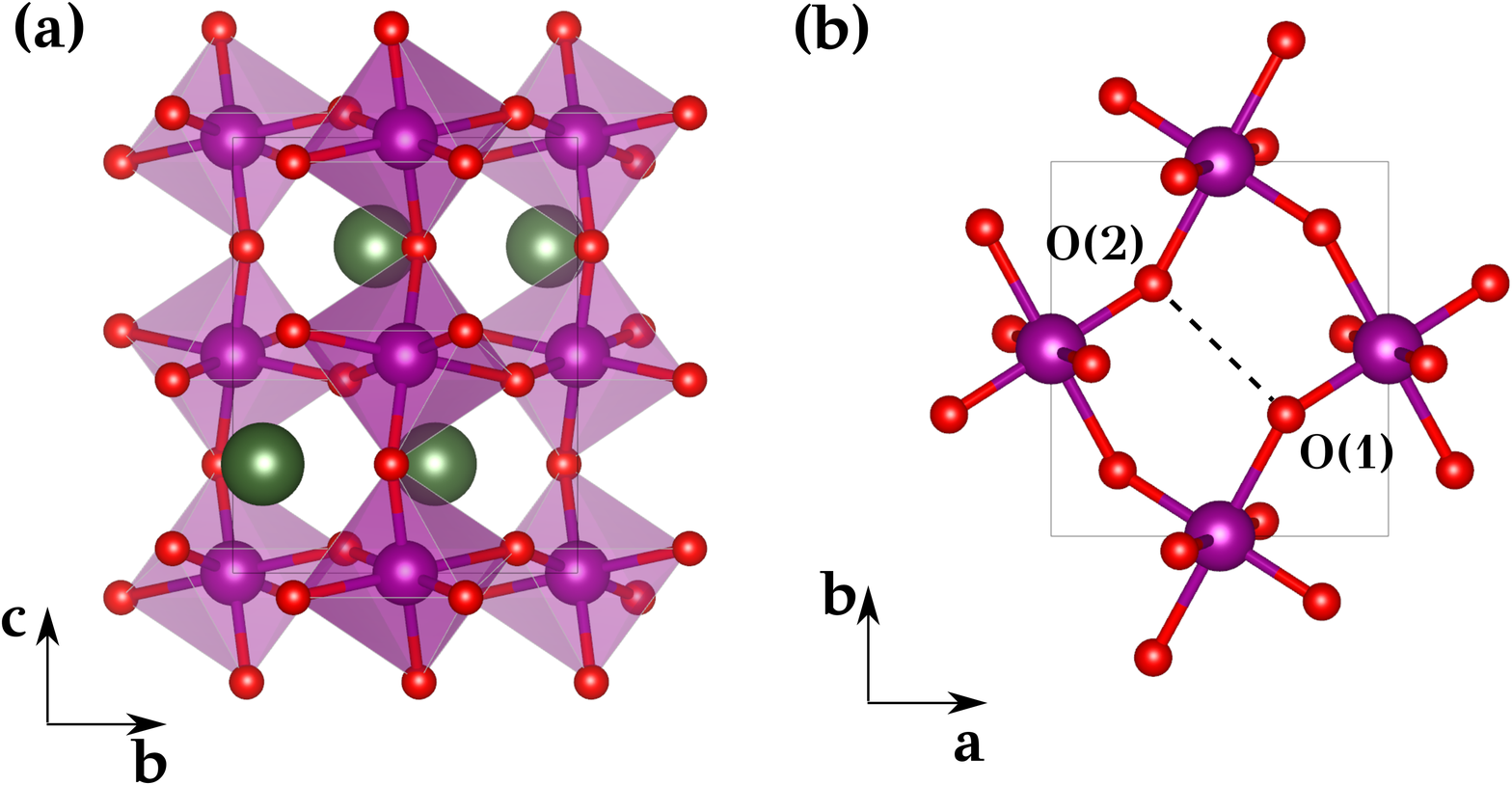}
\caption{\label{fig:rmno3_structure} Crystal structure of o-$R$MnO$_3$. Green spheres indicate $R^{3+}$ cations, purple - Mn$^{3+}$ cations and red - O$^{2-}$ anions. (a) shows the view in the $bc$ plane, (b) in the $ab$ plane (only Mn and O ions are shown).}
\end{figure}

In this work, we present a systematic study of the relationship between the crystal lattice and the magnetism in o-\textit{R}MnO$_3$ using X-ray diffraction measurements, density functional theory (DFT) and Monte Carlo (MC) simulations. We focus specifically on how the  microscopic exchange interactions can be varied by controlling the crystal lattice using chemical pressure or epitaxial strain and how these variations affect the long-range magnetic order. 
First, we employ non-resonant and resonant X-ray diffraction measurements to determine the lattice parameters and magnetic modulation vectors, respectively, in a set of epitaxially strained o-\textit{R}MnO$_3$ films. We show that the magnetic modulation vectors in highly strained films can differ significantly from those in bulk samples and relaxed films having the same \textit{R} cation. Next, we use DFT to calculate the internal coordinates for the experimentally reported crystal structures of several o-\textit{R}MnO$_3$ bulk samples (from the literature) and films (both from the literature and our new measurements) and analyze how the internal lattice parameters evolve across the series. We find that chemical pressure affects primarily the Mn-O-Mn bond angles, while epitaxial strain is accommodated by changes in the Mn-O bond lengths. To study the magnetism, we employ a model Hamiltonian which includes Heisenberg, biquadratic and four-spin ring exchange couplings, as well as Dzyaloshinskii-Moriya interactions (DMI) and single-ion anisotropy (SIA). We extract the exchange couplings and anisotropy constants by mapping the results of DFT calculations onto this model Hamiltonian and analyze how they are affected by the structural variations in bulk samples and films of o-\textit{R}MnO$_3$. We show that variations of the Mn-O-Mn bond angles caused by chemical pressure have a strong effect on the in-plane nearest-neighboring (NN) Heisenberg exchange while all the other couplings stay almost constant with changing \textit{R}. In turn, changes in the Mn-O bond lengths caused by epitaxial strain affect both in-plane and inter-plane NN Heisenberg couplings as well as next-nearest-neighboring (NNN) couplings and higher order exchanges. Then we use the calculated exchanges and anisotropies in a series of Monte Carlo simulations to determine the magnetic ground states and corresponding magnetic modulation vectors $q_b$. The latter are then compared to experimental values reported in the literature and obtained in this work through resonant X-ray diffraction measurements. We show that for most bulk and strained systems our model Hamiltonian and calculated couplings reproduce well the experimentally reported values of $q_b$. Moreover, we find that unconventional H-AFM and I-AFM orders can be stabilized in the strained films of o-LuMnO$_3$. Finally, we discuss the nature of the ferroelectricity that is induced in bulk and strained o-\textit{R}MnO$_3$ by the magnetic phases obtained in our MC simulations.

This article is organized as follows: In Sec. \ref{sec:structure} we describe the crystal structure and its relation to microscopic exchange interactions in o-\textit{R}MnO$_3$. In Sec.\ \ref{sec:phase_diagram} we introduce the magnetoelectric phase diagram of bulk o-\textit{R}MnO$_3$ and summarize the literature data on studies of the magnetic and ferroelectric properties of o-\textit{R}MnO$_3$ under hydrostatic pressure and epitaxial strain.  In Sec.\ \ref{sec:experiment}
we present the details of the experimental procedure and the results of our X-ray diffraction measurements. Then, in Sec.\ \ref{sec:computations} we introduce the magnetic model Hamiltonian which is used to describe the magnetism in o-\textit{R}MnO$_3$ and summarize the details of our DFT and MC simulations. In Sec.\ \ref{sec:results:structure} we present the results of our theoretical study of the evolution of internal lattice parameters in bulk and films of o-\textit{R}MnO$_3$. In Sec.\ \ref{sec:couplings} we present the calculated exchange couplings and anisotropies for all considered o-\textit{R}MnO$_3$ samples. In Sec.\ \ref{sec:montecarlo} we show which magnetic phases are stabilized in MC simulations using the calculated exchange coupling and anisotropy constants for the considered o-\textit{R}MnO$_3$ samples. In Sec.\ \ref{sec:polarization} we present the electric polarizations calculated for several representative bulk and strained o-\textit{R}MnO$_3$ using the magnetic phases obtained in our MC simulations.  Finally, in Sec.\ \ref{sec:summary} we summarize the main results of our investigation.

\section{Background and motivation}

\subsection{Crystal structure and exchange interactions in o-$R$MnO$_3$}
\label{sec:structure}
The o-\textit{R}MnO$_3$ have $Pbnm$ (\#62) symmetry and differ from the perfect cubic perovskites by the presence of Jahn-Teller (JT) distortions of the MnO$_6$ octahedra \cite{kanamori1960JT} and GdFeO$_3$-type (GFO) tiltings of these octahedra \cite{woodward1997tiltings} (see Fig.\ \ref{fig:rmno3_structure}). The JT distortions lift the degeneracy of the singly occupied majority spin $e_g$ states of the Mn$^{3+}$ ions (3$d^4$: $t^3_{2g}e^1_g$). The resulting occupied $e_g$ state on each Mn site $i$ can be represented as a linear combination of $\vert d_{z^2} \rangle$ and  $\vert d_{x^2-y^2} \rangle$ orbitals:
\begin{equation}
\label{eq:orb_mix_state}
\vert \phi_i \rangle=\mathrm{cos} \Big(\frac{\theta_i}{2} \Big) \vert d_{z^2} \rangle+\mathrm{sin} \big(\frac{\theta_i}{2} \Big) \vert d_{x^2-y^2} \rangle,
\end{equation}
where $\theta_i$ is the orbital mixing angle, which is determined by the balance between the energy of the orbital-lattice interaction and the elastic energy \cite{khomskii1973orbital_ordering}. A simple estimate of $\theta_i$ can be made using the following formula \cite{khomskii2005orbitals}:
\begin{equation}
\theta_i=\arctan\left(\frac{\sqrt[]{3}\left(l-s\right)}{2m-l-s}\right),
\label{eq:orb_mixing}
\end{equation}
where $s$, $m$ and $l$ are the lengths of short, medium and long Mn-O bonds in the MnO$_6$ octahedron.
The cooperative character of the JT distortions leads to an ordering of the occupied $e_g$ orbitals with $\theta_i=-\theta_j$ on the NN Mn sites $i$ and $j$ within the $ab$ planes, and $\theta_i=\theta_j$ along the $c$ direction. The GFO distortion in o-\textit{R}MnO$_3$ reduces the unit cell volume and so is larger for \textit{R} cations with smaller radii. This distortion reduces the Mn-O-Mn bond angles and decreases the lengths of the O(1)-O(2) bridges within the $ab$ planes (see Fig.\ \ref{fig:rmno3_structure} (b)). 

\begin{figure}
\centering
\includegraphics[width=0.45\textwidth]{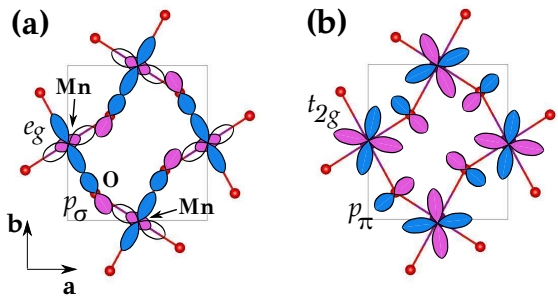}
\caption{\label{fig:orbitals} $d$-$p$-$d$ superexchange paths within the $ab$ planes in o-$R$MnO$_3$.
(a) $e_g$-$p_\sigma$-$e_g$ superexchange paths. The cooperative JT distortion of MnO$_6$ octahedra favors the ordering of the $e_g$ orbitals such that an occupied $e_g$ orbital (colored) on one Mn site overlaps with an empty $e_g$ orbital (white) on the neighboring Mn site via the $p_\sigma$ state of oxygen. (b) $t_{2g}$-$p_\pi$-$t_{2g}$ superexchange paths.}
\end{figure}

The magnetic ground state in o-\textit{R}MnO$_3$ is defined by the network of competing exchange couplings between the spins on NN and NNN Mn sites. Each superexchange interaction contains contributions from both $e_g$ and $t_{2g}$ orbitals mediated by the $p$ states of O anions. 
The crystal structure plays an important role in defining the relative strength of NN and NNN exchange couplings as well as the contributions from $e_g$ and $t_{2g}$ states to each coupling.
Indeed, for interactions within the $ab$ planes, the presence of the $e_g$ orbital ordering described above 
leads to superexchange between an occupied $e_g$ orbital on one Mn site with an empty $e_g$ orbital on the NN Mn site through the $p_\sigma$ states of O (see Fig.\ \ref{fig:orbitals} (a)). This favors ferromagnetic (FM) coupling between the $e_g$ spins according to the Goodenough-Kanamori-Anderson (GKA) rules \cite{goodenough1955superexchange,kanamori1959superexchange,anderson1959superexchange}. The $t_{2g}$ states, in turn, form covalent bonds with the $p_\pi$ states of O anions (see Fig.\ \ref{fig:orbitals} (b)) and electron transfer along the path $t_{2g}$-$p_\pi$-$t_{2g}$ favors antiferromagnetic (AFM) coupling of the $t_{2g}$ spins. Thus $e_g$ and $t_{2g}$ contributions compete with each other within the $ab$ planes. In general, the  $e_g$-$p_\sigma$-$e_g$ contribution is expected to be larger than that of the $t_{2g}$-$p_\pi$-$t_{2g}$ in absolute values, because the $e_g$ states are directed towards the $p_\sigma$ states of O, which provides a stronger overlap between them and, therefore, stronger coupling. Thus the resulting NN exchange within the $ab$ planes is expected to be FM. Nevertheless, the relative strengths of $e_g$-$p_\sigma$-$e_g$ and $t_{2g}$-$p_\pi$-$t_{2g}$ contributions can be changed by varying the bond angles and bond lengths (the amplitudes of the GFO and JT distortions, respectively), which can be achieved by hydrostatic or chemical pressure, or epitaxial strain. 
In fact, the change in the Mn-O bond lengths affects the overlap integral between the orbitals participating in the superexchange, and should modify both $e_g$ and $t_{2g}$ contributions by decreasing as bond lengths increase and vice versa. One has to keep in mind that variation of the Mn-O bond lengths can also change the mixing of the two $e_g$ states on each Mn site (in other words, the orbital mixing angle), which can in turn affect the $e_g$-$p_\sigma$-$e_g$ interaction. The variation of Mn-O-Mn bond angles is expected to influence the $e_g$-$p_\sigma$-$e_g$ coupling significantly due to the geometry of this bond (the coupling decreases with reducing angle and vice versa), while the $t_{2g}$-$p_\pi$-$t_{2g}$ should be less affected due to the isotropic character of $t_{2g}$ orbitals within the $ab$ planes. Changes in the GFO distortion can also modify the NNN exchange interactions along the $b$ direction due to variation of the lengths of the O(1)-O(2) bridges (see Fig.\ \ref{fig:rmno3_structure} (b)). Indeed, an increasing GFO distortion brings oxygens O(1) and O(2) closer to each other, which enhances the hybridization between their $p$ orbitals and leads to stronger coupling.  Along the $c$ direction the interactions occur between empty $e_g$ states mediated by $p_\sigma$ orbitals and between singly occupied $t_{2g}$ states mediated by $p_\pi$ orbitals; both are antiferromagnetic according to the GKA rules. Therefore, in this case $e_g$ and $t_{2g}$ contributions reinforce each other \cite{zhou2006rmno3}.

\subsection{Experimental phase diagram of bulk o-\textit{R}MnO$_3$}
\label{sec:phase_diagram}
\begin{figure}
\centering
\includegraphics[width=0.48\textwidth]{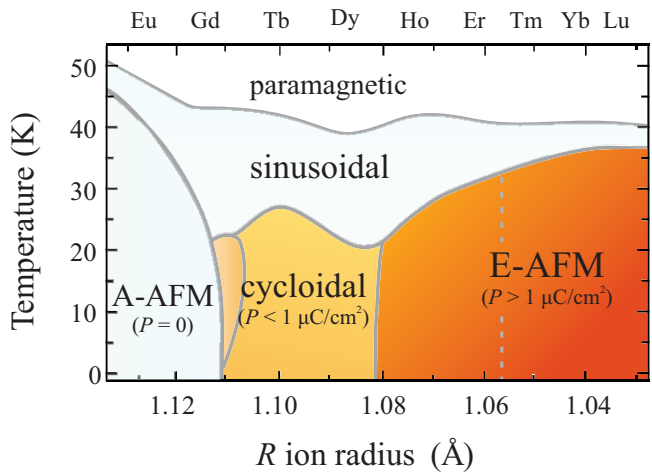}
\caption{\label{fig:PD_2_options} Phase diagram of bulk o-\textit{R}MnO$_3$, following Ref.\ \onlinecite{ishiwata2010perovskite}. Borders are drawn based on magnetic susceptibility and electric polarization measurements conducted on powders. Labels at the top of the image indicate the \textit{R} ion associated with the radii indicated on the lower horizontal axis. Increased orange shading indicates higher predicted electric polarization.}
\end{figure}

\begin{figure}
\centering
\includegraphics[width=0.48\textwidth]{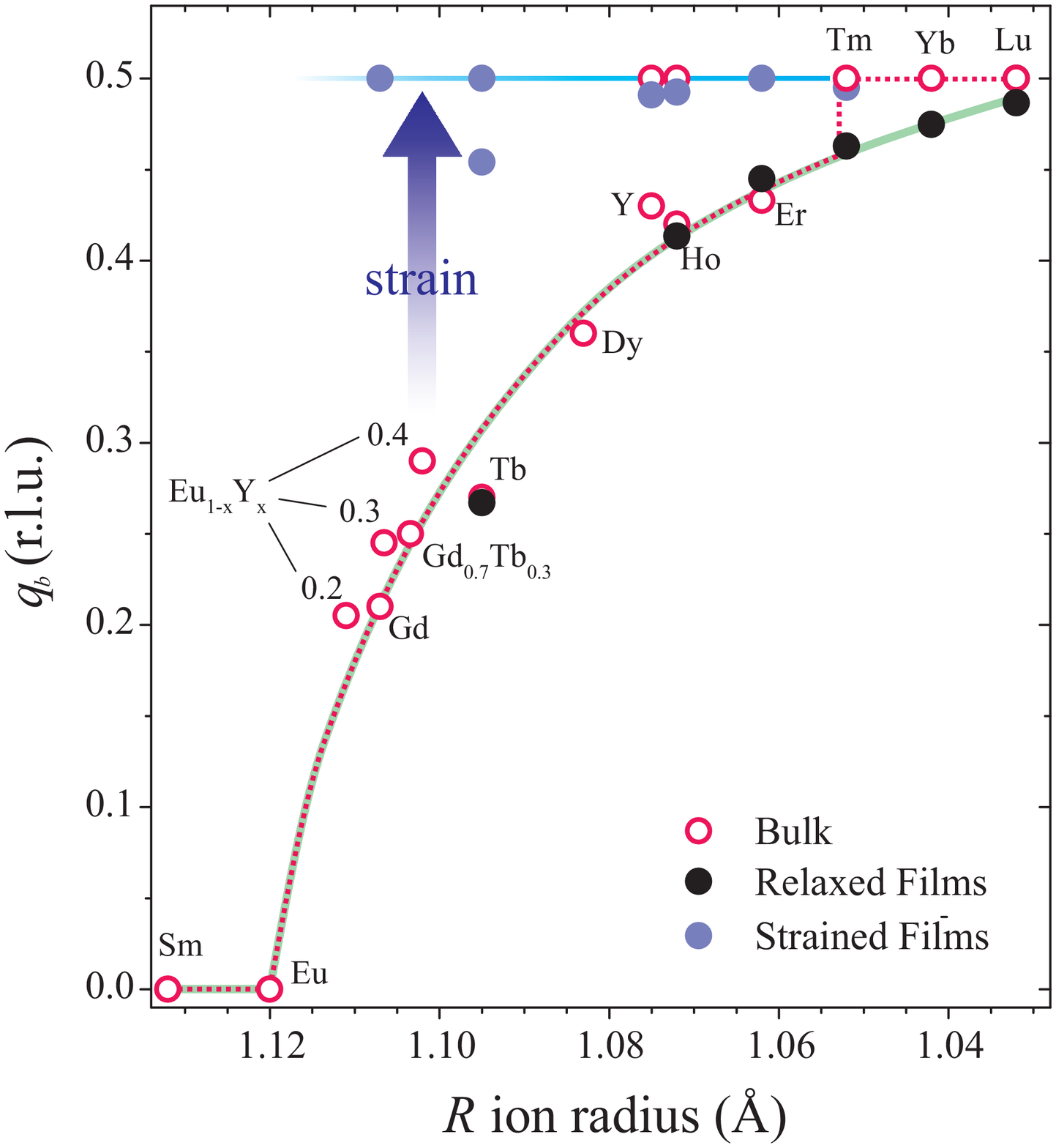}
\caption{\label{fig:q_Vs_R} Magnetic modulation vector \textit{q$_b$} as function of radius of the \textit{R} ion. Literature data for bulk powder samples of o-\textit{R}MnO$_3$ are presented as empty circles. Films (literature data from Refs. \onlinecite{wadati2012origin,windsor2014multiferroic,windsor2015interplay} and our new measurements) are indicated by filled circles. The dashed line indicates the trend for bulk materials, and the green solid line that for relaxed films. The blue line shows the discrepancy between bulk samples and strained films with the same $R$ ion. $q_b$ is in reciprocal lattice units.}
\end{figure}

The interplay between lattice and spin degrees of freedom described in Sec.\ \ref{sec:structure} manifests in the magnetic phase diagram of bulk o-\textit{R}MnO$_3$ (see Fig.\ \ref{fig:PD_2_options}), which was experimentally established through multiple studies of magnetic order in these materials. At low temperatures, o-\textit{R}MnO$_3$ with larger $R$ ion species ($R$=La,...,Gd) exhibit an A-type AFM ordered ground state (A-AFM, modulation vector $q_b=0$) \cite{Wollan1955ABC} which is favored by their orbital ordering. Decreasing the radius of the $R$ cation increases the GFO distortion, which leads to an evolution of the magnetic order, initially to incommensurate (IC) spiral structures ($R$=Tb, Dy) and then to E-type AFM order (E-AFM, $q_b=\nicefrac{1}{2}$; $R$=Tm, Yb and Lu). Conflicting reports exist for the intermediate radii of Ho and Y. For o-HoMnO$_3$ both E-AFM order \cite{munoz2001homno3} and incommensurate order with $q_b\approx0.4$ \cite{brinks2001crystal}, identified as a sinusoidal spin density wave, have been reported. For o-YMnO$_3$ an IC \textit{ac} spiral ($q_b=0.078$) and sinusoidal spin density wave ($q_b=0.435$) \cite{munoz2002ymno3} have both been observed, and E-AFM order has been reported based on a study of the structural modulation at low temperatures.  Lastly, an incommensurate magnetic structure has also been reported for o-ErMnO$_3$ \cite{ye2007incommensurate}, with a similar propagation vector ($q_b=0.433$) to that of o-HoMnO$_3$ and o-YMnO$_3$, but the magnetic structure was not specified. In Ref.\ \onlinecite{ishiwata2010perovskite} this phase was discussed in terms of coexisting spiral and E-AFM phases. In our recent theoretical study based on DFT and Monte Carlo simulations for o-HoMnO$_3$ and o-ErMnO$_3$, we demonstrated that this IC magnetic phase is likely a "w-spiral" order \cite{fedorova2018fourspin}. Since all the aforementioned magnetic phases can be described by the modulation vector $\mathbf{q}$=$(0,q_b,1)$, the evolution of magnetism across the series of bulk o-$R$MnO$_3$ can be represented as a variation of $q_b$ with decreasing radius of the $R$ cation ($r_R$). In Fig.\ \ref{fig:q_Vs_R} 
we summarize the literature values of the modulation vectors $q_b$ reported for bulk o-$R$MnO$_3$ (single crystals and powders) \cite{brinks2001crystal,munoz2002ymno3,ye2007incommensurate,pomjakushin2009evidence,okamoto2008neutron,yamasaki2007mixtures,Yamasaki2008mixtures,OFlynn2011SmMnO3} shown as red open circles versus $r_R$. One can see that \textit{q$_b$} varies systematically with $r_R$, from $q_b=0$ for the A-AFM phase to $q_b=\nicefrac{1}{2}$ for the E-AFM phase. The two contradicting values for $R$ = Ho and Y are also presented.

\begin{figure}
\centering
\includegraphics[width=0.43\textwidth]{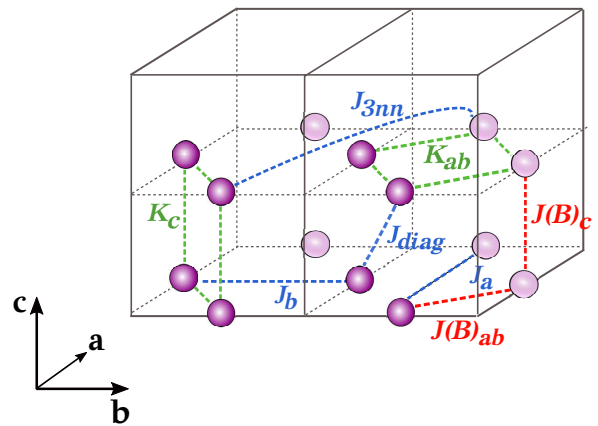}
\caption{\label{fig:exchanges} Heisenberg ($J_c$, $J_{ab}$, $J_a$, $J_b$, $J_{diag}$, $J_{3nn}$), biquadratic ($B_c$ and $B_{ab}$) and four-spin ring ($K_c$ and $K_{ab}$) exchange interactions considered in the model Hamiltonian of Eq.\ \ref{eq:fullHam}. The 40 atom o-$R$MnO$_3$ supercell (1$\times$2$\times$1 of the 20-atom unit cell) containing 8 Mn ions (purple spheres, the lighter spheres indicate Mn ions in neighboring cells) is shown ($R$ and O ions are not shown).}
\end{figure}

The microscopic mechanism that drives this evolution of magnetism in bulk o-\textit{R}MnO$_3$ is still debated. It was initially considered in terms of competing Heisenberg exchange interactions between NN and NNN Mn$^{3+}$ spins within the $ab$ planes: the relative strengths of these interactions are directly affected by the increasing GFO distortion (decreasing Mn-O-Mn bond angles) \cite{zhou2006rmno3,kimura2003distorted}. Ref.\ \onlinecite{mochizuki2011theory} suggested that an increase of the GFO distortion primarily affects the NNN in-plane coupling $J_b$ (see Fig.\ \ref{fig:exchanges}), and that this causes the evolution of magnetism. However, recent theoretical studies have demonstrated that this effect mainly reduces the NN coupling $J_{ab}$, and that magnetic order evolves because the effect of other couplings (such as NNN Heisenberg and higher order exchanges) becomes more pronounced \cite{fedorova2015biquadratic,zhang2018rmno3}.  
In Ref.\ \onlinecite{solovyev2009j3nn} the importance of the interaction between third-nearest-neighboring Mn spins within the $ab$ planes ($J_{3nn}$ in Fig.\ \ref{fig:exchanges}) was demonstrated. Moreover, it was shown that biquadratic exchange interactions play a key role in stabilizing E-AFM over spiral order \cite{kaplan2009biquadratic,hayden2010biquadratic}. We recently found that inter-plane four-spin ring exchange interactions ($K_c$ in Fig. \ \ref{fig:exchanges}) are crucial to explain the establishment of the w-spiral state in o-HoMnO$_3$ and o-ErMnO$_3$, as well as two unconventional commensurate magnetic phases (so-called H-AFM and I-AFM) which can, in principle, form in these systems \cite{fedorova2018fourspin}. 

Understanding the interplay between ferroelectricity and magnetism has been a central motivator for studying this family of materials. IC spiral and E-AFM orders
break inversion symmetry and induce an electric polarization in o-$R$MnO$_3$, making them type II multiferroics. For the 
spiral orders, the electric polarization is usually treated as an effect arising from spin-orbit coupling and is explained in terms of the spin-current model \cite{katsuraPRLknb_model} and/or  antisymmetric exchange striction  \cite{sergienko2006DMI}. 
Since spin-orbit coupling is weak in o-$R$MnO$_3$, the resulting electric polarization is relatively small ($P\approx0.1$ $\mu$C/cm$^2$) \cite{kimura2003magnetic,kimura2005magnetoelectric}. In systems with E-AFM order, the proposed mechanism for the magnetically induced electric polarization is symmetric exchange striction.
It was theoretically predicted by Sergienko \textit{et al.} \cite{sergienko2006ferroelectricity} that this mechanism should provide significantly enhanced polarization values ($P\approx0.5-12$ $\mu$C/cm$^2$) compared to systems with spiral order. This prediction was confirmed later by Berry phase calculations for o-HoMnO$_3$ ($P\approx 6$ $\mu$C/cm$^2$) \cite{picozzi2007dual}. 
However, to the best of our knowledge, the predicted polarization values for E-AFM order have not been experimentally detected in bulk o-$R$MnO$_3$; the largest $P$ values were reported for o-LuMnO$_3$ and o-YMnO$_3$, reaching 0.17 and 0.24 $\mu$C/cm$^2$, respectively \cite{chai2012rmno3_polar,okuyama2011magnetically}, which is at least an order of magnitude smaller than $P$ obtained from first principles. This contradiction between theory and experiment is still not fully understood. Moreover, measurements of $P$ in o-$R$MnO$_3$ with $R$=Ho, Er and Y gave contradictory results. 
For example, in Ref. \onlinecite{lorenz2007homno3_ymno3} $P \approx 0.009$ $\mu$C/cm$^2$ was observed in o-HoMnO$_3$ along the $a$ axis, while Ref.\ \onlinecite{lee2011mechanism} reported $P \approx 0.15$ $\mu$C/cm$^2$ along the $c$ direction. In both cases the importance of the Ho$^{3+}$ $f$-electron moments in inducing $P$ was underlined, since $P$ demonstrated a drastic increase only below their ordering temperature. Ref. \onlinecite{lorenz2007homno3_ymno3} reported $P \approx 0.025$ $\mu$C/cm$^2$ for o-YMnO$_3$. The origin of this value is not yet understood since neither the reported sinusoidal spin density wave nor the $ac$ spiral are expected to produce an electric polarization according to the mechanisms of magnetically induced ferroelectricity, described above. This also cannot be explained by an ordering of $R^{3+}$ moments as Y has an empty $f$-shell. In o-ErMnO$_3$ no sizable polarization was measured by Ye \textit{et al.} \cite{ye2007incommensurate}, while Ishiwata \textit{et al.} reported $P\approx0.06$ $\mu$C/cm$^2$ for this compound \cite{ishiwata2010perovskite}. 

Recent experimental studies have demonstrated that structural modifications due to hydrostatic pressure or epitaxial strain can stabilize magnetic phases in o-$R$MnO$_3$ that are different from those that are stable in unperturbed bulk samples. Moreover, the electric polarization in such structurally modified samples can be significantly larger than in bulk samples. For example, the magnetic order in o-TbMnO$_3$ evolves under high pressure from a spiral to the E-AFM state \cite{makarova2011pressure,aoyama2014tbmno3} which produces $P\approx1$ $\mu$C/cm$^2$. It was recently demonstrated that films of o-$R$MnO$_3$ with $R$=Gd,...,Lu epitaxially grown on YAlO$_3$ yield electric polarizations of up to 1 $\mu$C/cm$^2$ (for o-TbMnO$_3$ $P$ of up to 2 $\mu$C/cm$^2$ was measured), suggesting that the E-AFM phase is likely stabilized \cite{shimamoto2017phase_diagram}. Strain was also found to affect or even tune the magnetic modulation vector \cite{windsor2014multiferroic}. This effect was detailed in a recent study on o-HoMnO$_3$ films. A strained film of o-HoMnO$_3$ (32 nm [010]-oriented film grown on YAlO$_3$ substrate) was shown to possess a magnetic modulation vector of $q_b\approx0.49$, while a relaxed film (120 nm) had $q_b\approx0.42$, which is close to that of bulk o-HoMnO$_3$ \cite{brinks2001crystal}. Both films showed enlarged polarization values compared to bulk o-HoMnO$_3$ \cite{shimamoto2016HoMnO3}. 
In spite of these advances in structural manipulation, the underlying mechanism behind the evolution of magnetic order in the strained samples as well as the enhancement of the polarization remain to be understood. 

From all the literature data summarized above, it is clear that both magnetism and ferroelectricity in o-$R$MnO$_3$ can be manipulated by structural variations, such as hydrostatic or chemical pressure or epitaxial strain. Since the ferroelectricity in these materials is governed by magnetism, an understanding of the relationship between the crystal lattice and magnetic orders is of primary importance for potential optimization of their multiferroic properties.  

\section{Experiments on crystalline films}
\label{sec:experiment}

\begin{figure}
\centering
\includegraphics[trim={0cm 2.5cm 0cm 0cm},width=0.6\textwidth]{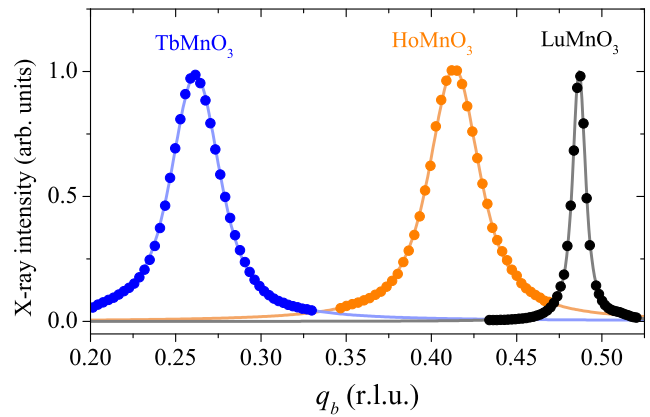}
\caption{\label{fig:raw_data} Magnetic intensity from reciprocal space scans along the [010] direction, taken using resonant X-ray diffraction at the Mn $L_3$ edge. Data are from [010]-oriented films of orthorhombic TbMnO$_3$, HoMnO$_3$ and LuMnO$_3$. $q_b$ is in reciprocal lattice units.}
\end{figure}

As a basis for studying the effects of epitaxial strain on the relationship between the lattice and magnetic order, we 
measured the lattice parameters and magnetic modulation vectors of a selection of epitaxially grown films. These were grown by pulsed laser deposition using stoichiometric ceramic targets of the corresponding hexagonal \textit{R}MnO$_3$ materials. Further growth details are found in Ref.\ \onlinecite{windsor2014multiferroic}. A full list of films discussed here is available in Table I of the Supplemental materials. 

Non-resonant X-ray diffraction (XRD) was employed to measure lattice constants to high precision using the Surface Diffraction end station of the Materials Science beam line of the Swiss Light Source (SLS) \cite{Willmott2013MS}. The lattice constants were determined by collecting precise motor positions of several reflections and computing the best fit to a UB matrix of an orthorhombic crystal. The photon energies used were all between 8 and 10 keV. Diffracted intensities were collected using a Pilatus 100K detector \cite{Broennimann2006Pilatus} mounted on the detector arm. In both experiments samples were mounted on the cold head of a Janis flow cryostat. The measured lattice parameters for all considered o-\textit{R}MnO$_3$ films are summarized in Table I of the Supplemental materials.

Resonant X-ray diffraction (RXD) experiments were conducted to probe antiferromagnetic order. These were done using the RESOXS UHV diffraction end station \cite{Staub2008RESOXS} at the SIM beam line \cite{Flechsig2010SIM} of the SLS. Photon energies used correspond to the Mn $L_3$ absorption edge using $\pi$-polarized incident light (electric field in the scattering plane). Data were taken at 10 K. Scattered intensities were collected using an IRD AXUV100 photodiode. Scans were conducted along the [010] direction of reciprocal space, following the (0,$q_b$,0) magnetic reflection. This reflection provides a direct and unequivocal measure of the modulation parameter $q_b$. 
In Fig.\ \ref{fig:raw_data} we present as an example the scans for [010]-oriented o-TbMnO$_3$ (150 nm), o-HoMnO$_3$ (120 nm) and o-LuMnO$_3$ (104 nm) films. 

In Figure \ref{fig:q_Vs_R} we present our measured $q_b$ values for o-$R$MnO$_3$ films with different $R$ ions and different levels of strain (see details in Table I of Supplemental materials) alongside the literature values for bulk samples and additional literature values for films. Two notable observations can be made. First, despite having the same $R$ ion, relaxed films follow the gradual trend of the bulk samples, while highly strained films tend towards locking to the commensurate $q_b = \nicefrac{1}{2}$ value.
Second, for the lower $r_R$ values (Tm, Yb, and Lu), the bulk o-$R$MnO$_3$ samples have $q_b = \nicefrac{1}{2}$, but relaxed films do not 
reach this value, and instead show a gradual 
evolution of $q_b$.
These film-bulk discrepancies support the idea that small variations in the crystal lattice have a strong effect on
the position of a material in the magnetic phase diagram and serve as a motivation for our theoretical study of the relationship between $q_b$ and the crystal lattice in o-$R$MnO$_3$. 

\section{Computational details}
\label{sec:computations}
\subsection{Spin model Hamiltonian}
\label{sec:hamiltonian}

In order to accurately describe the complex magnetic phase diagram of the o-$R$MnO$_3$ series (see Sec.\ \ref{sec:phase_diagram}), we employ the following spin model Hamiltonian:
\begin{equation}
\label{eq:fullHam}
H=H_{Heis}+H_{BQ}+H_{4sp}+H_{SIA}+H_{DM},
\end{equation}
where 
\begin{equation}
H_{Heis}=\sum_{<i,j>}J_{ij}(\mathbf{S}_i\cdot\mathbf{S}_j), 
\label{eq:HeisHam}
\end{equation}
\begin{equation}
H_{BQ}=\sum_{<i,j>}B_{ij}(\mathbf{S}_i\cdot\mathbf{S}_j)^2,
\label{eq:biqHam}
\end{equation}
\begin{eqnarray}
H_{4sp}=\sum_{<i,j,k,l>}K_{ijkl}\left[ \left(\mathbf {S}_i\cdot\mathbf {S}_j\right)\left(\mathbf S_k\cdot\mathbf S_l\right) \right. \nonumber \\ + \left. \left(\mathbf S_i\cdot\mathbf S_l\right)\left(\mathbf S_k\cdot\mathbf S_j\right) - \left(\mathbf S_i\cdot\mathbf S_k\right)\left(\mathbf S_j\cdot\mathbf S_l\right)\right], 
\label{eq:4bodyHam}
\end{eqnarray}
\begin{equation}
H_{SIA}=A\sum_{i}S^2_{i,b}, 
\label{eq:siaHam}
\end{equation}
\begin{equation}
H_{DM}=\sum_{<i,j>}\mathbf{D}_{ij}\cdot[\mathbf{S}_i\times\mathbf{S}_j].
\label{eq:dmHam}
\end{equation}

The first term, $H_{Heis}$ (Eq.\ \ref{eq:HeisHam}), is a Heisenberg Hamiltonian, where $J_{ij}$ are exchange interactions between spins $\mathbf{S}_i$ and $\mathbf{S}_j$ on Mn sites $i$ and $j$, respectively. A $H_{Heis}$ including only AFM $J_c$ and $J_{b}$ and FM $J_{ab}$ (see Fig.\ \ref{fig:exchanges}) can explain the establishment of the A-AFM and spiral orders, the latter occuring if the NNN $J_b$ is large enough to compete with NN $J_{ab}$. 
We extend our model by including also the second NN couplings along the $c$ direction ($J_{diag}$, see Fig.\ \ref{fig:exchanges}) and second ($J_a$) and third NN exchanges ($J_{3nn}$) within the $ab$ planes. 
Further neighbor couplings are not taken into account since we showed in our previous work that they are negligible in comparison with those mentioned above \cite{fedorova2015biquadratic}. 

The second term, $H_{BQ}$ (Eq.\ \ref{eq:biqHam}), describes the biquadratic exchange interactions between spins $\mathbf{S}_i$ and $\mathbf{S}_j$. It has been demonstrated that the biquadratic couplings between NN spins within the $ab$ planes, $B_{ab}$, are crucial for establishment of E-AFM order \cite{kaplan2009biquadratic,hayden2010biquadratic}.  In this work we consider NN biquadratic couplings, both within the $ab$ planes ($B_{ab}$) and along the $c$ direction ($B_c$) (see Fig.\ \ref{fig:exchanges}). 

The third term, $H_{4sp}$ (Eq. \ref{eq:4bodyHam}), corresponds to the four-spin ring exchange couplings, which arise from consecutive electron hoppings between the NN Mn ions forming four-site plaquettes. We recently showed that the energies of different magnetic orders calculated using DFT for several o-\textit{R}MnO$_3$ cannot be accurately fitted to the isotropic spin Hamiltonian including only Heisenberg and biquadratic exchanges, and the four-spin ring terms need to be included to provide an accurate description of the magnetism \cite{fedorova2015biquadratic}. Moreover, we found that the presence of strong inter-plane four-spin ring exchange $K_c$ can stabilize several exotic magnetic orders in o-\textit{R}MnO$_3$  such as incommensurate w-spiral and commensurate H-AFM and I-AFM (see Ref.\ \onlinecite{fedorova2018fourspin} for details). Here we include in the analysis the four-spin interactions in two types of plaquettes: those within the $ab$ planes ($K_{ab}$) as well as inter-plane ($K_c$) plaquettes (Fig.\ \ref{fig:exchanges}).

The fourth term, $H_{SIA}$ (Eq.\ \ref{eq:siaHam}), is a single ion anisotropy which sets the magnetic easy axis along the $b$ direction. 

The fifth term, $H_{DM}$ (Eq.\ \ref{eq:dmHam}), describes the Dzyaloshinskii-Moriya interactions. We consider DM vectors, $\mathbf{D}_{ij}$, which are defined both for Mn-O-Mn bonds along the $c$ direction and within the $ab$ planes \cite{mochizuki2009microscopic,solovyev1996lamno3}. As shown in Ref.\ \onlinecite{solovyev1996lamno3}, due to the symmetry of o-$R$MnO$_3$ crystals, their DM vectors can be described using five parameters: $\alpha_{ab}$, $\beta_{ab}$ and $\gamma_{ab}$ for the in-plane DM interactions ($\mathbf{D}_{ij}^{ab}$) and $\alpha_{c}$ and $\beta_c$ for the inter-plane ones ($\mathbf{D}_{ij}^c$) (see Fig.\ 3 in Ref. \onlinecite{solovyev1996lamno3}). The $\alpha_c$ components of the $\mathbf{D}_{ij}^c$ vectors favor a canting of the Mn spins from the $b$ axis towards the $c$ axis \cite{mochizuki2009microscopic,solovyev1996lamno3}, which was experimentally observed for several o-\textit{R}MnO$_3$  \cite{matsumoto1970lamno3, mukherjee2017lumno3}. The $\gamma_{ab}$ components of the $\mathbf{D}_{ij}^{ab}$ vectors can favor stabilization of the $ab$ spiral instead of the $bc$ spiral \cite{mochizuki2009microscopic}. 
In this work we consider only $\alpha_{c}$ and $\gamma_{ab}$ and neglect all other components of the DM vectors.

\subsection{First-principles calculations}
All density functional calculations are performed using the Vienna \textit{Ab initio} Simulation Package (VASP) based on the projector-augmented plane wave (PAW) method of DFT \cite{kresseVasp}. 
We employ the generalized gradient approximation with Hubbard $U$ correction (GGA+$U$) for the exchange-correlation potential in the form of Perdew, Burke and Ernzerhof (PBE) revised for solids (PBEsol)\cite{perdew2008Pbesol} as it gives better agreement between theoretically optimized and experimental lattice parameters for the considered systems in comparison with the standard PBE \cite{perdew1996pbe}. The parameter of the on-site Coulomb repulsion for the Mn $d$ states is set to $U$=1 eV and the on-site exchange interaction to $J_H$=0 eV since these values give reasonable sizes of the band gaps and correct magnetic ground states for many o-\textit{R}MnO$_3$.
The $f$ states of the rare-earth elements are treated as core states. The cutoff energy for the plane wave basis set is 600 eV. All the calculations using the 20 atom unit cells (structural relaxations, calculations of the biquadratic couplings, DMI and anisotropy constants) are performed with a $\Gamma$-centered 7$\times$7$\times$5 k-point mesh. For the 80 atom (2$\times$2$\times$1) supercells (calculations of the Heisenberg and four-spin ring exchanges) we use a $\Gamma$-centered 3$\times$3$\times$5 k-point mesh and for 80-atom 1$\times$2$\times$2 supercells (calculations of electric polarizations) we use a $\Gamma$-centered 7$\times$3$\times$2 k-point mesh. For the lattice optimizations the structures are considered to be relaxed if the Hellmann-Feynman forces acting on the atoms are smaller than $10^{-4}$ eV/{\AA} and, when the volume is allowed to relax, the components of the stress tensor are smaller than 0.1 kbar. All the structural relaxations are performed with A-AFM order imposed. Spin-orbit coupling is included only in the calculations of the DMI and SIA.

\subsection{Monte Carlo simulations}
Monte Carlo simulations performed in this work are based on the Metropolis algorithm \cite{metropolis1953montecarlo} combined with overrelaxation moves \cite{creutz1987overrelaxation}. We employ the replica exchange technique \cite{swendsen1986replicas,earl2005partemp} which is efficient in finding a global energy minimum in systems with many local energy minima, which is the case for frustrated spin systems with many competing interactions. For each compound we simulate in parallel $M$=200 replicas, each at a different temperature. The range of temperatures is defined as $T_k$=$T_0/\alpha^k$, where $T_0$=0.005 meV is the temperature of interest, $k$=1...$M-1$ and $\alpha$=0.962 (this value gives the maximal temperature $T_{M-1}$ larger than the strongest exchange interactions in the considered systems). We consider unit cells containing two Mn atoms (in the following we call this the MC unit cell) - Mn$_1$ (0,0.5,0) and Mn$_2$ (0.5,1,0) - and perform simulations for different system sizes (12$\times$40$\times$12  and 4$\times$100$\times$4 MC unit cells). We apply periodic boundary conditions in all directions and double check the results using open boundary conditions along the $b$ direction (and periodic along the $a$ and $c$ axes) to ensure that the modulation vectors of the obtained magnetic structures are not affected by the choice of boundary conditions. We also perform calculations starting from different types of magnetic order - A-AFM, E-AFM, H-AFM (see our recent work, Ref.\ \onlinecite{fedorova2018fourspin}, for the details about the latter state) and random orientation - as an additional check that the results are not affected by the starting configurations and the systems are not trapped in a local energy minimum.
    
\section{Calculated effects of chemical pressure and epitaxial strain on the crystal lattice}
\label{sec:results:structure}
\begin{figure*}
\includegraphics[width=0.95\linewidth,trim=0cm 0cm 0cm 0cm]{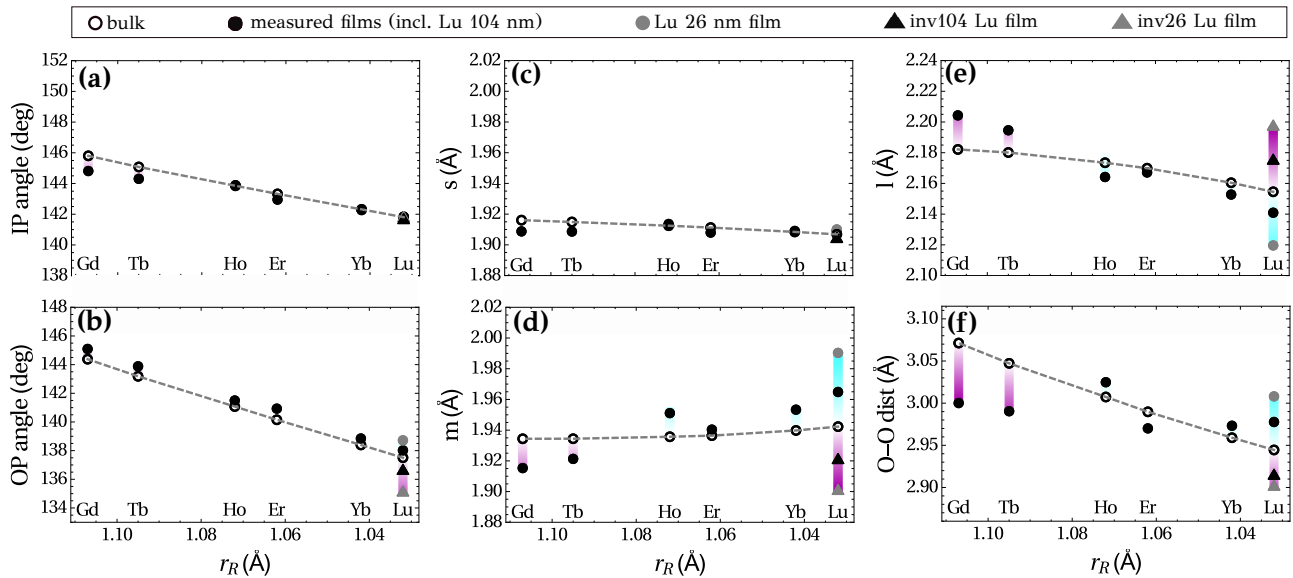} \\
\caption{\label{fig:str_vs_R}Theoretically optimized structural parameters of strained films and bulk o-$R$MnO$_3$ versus the radius of the $R$ cation: (a) and (b) give the Mn-O-Mn bond angles within the $ab$ planes (IP angle) and along the $c$ direction (OP angle), respectively;  
(c), (d) and (e) show short ($s$), medium ($m$) and long ($l$) Mn-O bond lengths of the MnO$_6$ octahedra, respectively; (f) shows the lengths of the O(1)-O(2) bridges (see Fig.\ \ref{fig:rmno3_structure} (b)) within the $ab$ planes. Bulk samples are shown by empty circles, strained films by filled circles. For LuMnO$_3$ the triangles denote calculations for hypothetical films which are compressively strained in the $ac$ plane by the same amount but in the opposite direction as the experimentally measured tensile strained films. LuMnO$_3$ 26 nm film and the corresponding inverse case are highlighted in gray, 104 nm film and the inverse case in black.  Compressive strain within the $ac$ planes of the o-$R$MnO$_3$ films is shown by the violet color, tensile strain by the blue color. The dashed lines connecting the data points for bulk o-$R$MnO$_3$ are guides to the eye.}
\end{figure*}

First we calculate how the o-\textit{R}MnO$_3$ crystal structure evolves under chemical pressure and epitaxial strain.
We start by considering bulk o-\textit{R}MnO$_3$ and analyze how the internal lattice parameters vary with the radius of the \textit{R} cation.  
For this purpose we consider several representatives of the o-\textit{R}MnO$_3$ series (namely \textit{R}=Gd, Tb, Ho, Er, Yb and Lu) and fully optimize their lattice parameters and internal coordinates using DFT, with the experimentally reported structures as the starting point \cite{mori2002lnmno3,alonso2000evolution,munoz2001homno3,ye2007incommensurate,tachibana2007jahn,okamoto2008neutron}. This allows us to make a direct comparison between our findings for bulk samples and for strained films, for which the internal coordinates are not readily measurable. In Fig.\ \ref{fig:str_vs_R} we present the obtained lengths of the short ($s$), medium ($m$) and long ($l$) Mn-O bonds within the MnO$_6$ octahedra, the O(1)-O(2) distances (see Fig.\ \ref{fig:rmno3_structure} (b)) as well as the Mn-O-Mn bond angles within the $ab$ planes (IPA) and along the $c$ axis (OPA) versus the $R$ radius. The exact values for all the optimized lattice parameters together with the experimentally reported values are summarized in Table II of the Supplemental materials. From Fig.\ \ref{fig:str_vs_R} (a) and (b) one can see that in bulk o-$R$MnO$_3$ the volume reduction due to decrease in the radius of the $R$ cation is almost fully accommodated by reducing the Mn-O-Mn bond angles within the $ab$ planes and along the $c$ direction. As a secondary effect, the O(1)-O(2) distances also decrease as the $R$ radius decreases (Fig.\ \ref{fig:str_vs_R} (f)). In turn, the $s$ and $m$ Mn-O bond lengths (Figs.\ \ref{fig:str_vs_R} (c) and (d), respectively) are almost constant across the series of the bulk samples and $l$ bonds decrease slightly from Gd to Lu (Fig.\ \ref{fig:str_vs_R} (e)). This is in agreement with literature experimental data \cite{zhou2006rmno3} as well as with previous theoretical reports \cite{yamauchi2008rmno3,zhang2018rmno3}. 

In the next step, we investigate the effects of strain on the 
crystal structure of o-$R$MnO$_3$. For this we consider a set of [010]-oriented o-\textit{R}MnO$_3$ films (with the same \textit{R} as in the bulk samples described previously in this section) grown epitaxially on YAlO$_3$ substrates. The experimental lattice parameters of the o-GdMnO$_3$ and o-TbMnO$_3$ films are taken from Refs.\ \onlinecite{shimamoto2017phase_diagram,shimamoto2017tbmno3} and for the other films we use values measured in this work. The lattice mismatch between the film and the substrate results in either compressive or tensile strain in the $ac$ planes:   
For o-GdMnO$_3$ and o-TbMnO$_3$ films the $a$ and $c$ lattice constants are strongly compressed compared to the corresponding bulk values, which in turn leads to an increase in $b$; in o-YbMnO$_3$ and the two o-LuMnO$_3$ films (26 nm and 104 nm) the effect is opposite - $a$ and $c$ are increased and $b$ is reduced; in o-ErMnO$_3$ the $a$ lattice constant is compressed, while $b$ and $c$ are increased. For comparison we also consider a o-HoMnO$_3$ film grown on a NdGaO$_3$ substrate, for which the $a$ and $c$ lattice constants of the film are extended and $b$ is significantly reduced. To simulate the epitaxially strained films, we constrain the lengths $l_i^{str}$ ($i=a,c$) of the $a$ and $c$ lattice constants to the values:
\begin{equation}
l_i^{str}=(1+\epsilon_i)l_i^{bulk},
\end{equation}
where $l_{i}^{bulk}$ is the corresponding lattice parameter of the relaxed bulk crystal structure described above and $\epsilon_i$ is the experimental strain applied to the $i$th lattice constant. Then we use DFT to optimize the length of the $b$ lattice parameter, which is perpendicular to the substrate, and the ionic positions. In Fig. \ref{fig:str_vs_R} we present the Mn-O-Mn bond angles, Mn-O bond lengths and O(1)-O(2) distances of the optimized strained crystal structures (together with the corresponding parameters for the bulk structures) versus the radius of the $R$ cation (all lattice parameter values are summarized in Table III of the Supplemental materials).  When we compare each bulk sample with its corresponding strained film(s), we see that applying strain (both compressive and tensile) affects mostly the $m$ and $l$ Mn-O bonds of the MnO$_6$ octahedra while the $s$ bonds as well as both IP and OP Mn-O-Mn bond angles remain almost unchanged between bulk and strained samples. For $m$ and $l$ bonds, compressive and tensile strains clearly have opposite effects: in the first case ($R$=Gd, Tb) $m$ is reduced and $l$ is increased (due to increase in the $b$ lattice constant), and vice versa in the latter case ($R$=Yb, Lu). 

Next, to check whether the effect of compressive strain can be different in systems with small unit cell volume, we simulated two hypothetical films of o-LuMnO$_3$ in which we artificially compressed the $a$ and $c$ axes of the fully optimized bulk crystal structure by the same amount as they expanded in the experimentally studied tensile strained o-LuMnO$_3$ films (26 nm and 104 nm)  described above. These hypothetical films will be called inv26 and inv104, respectively, in the following. The obtained lattice parameters for the inv26 and inv104 o-LuMnO$_3$ films are also shown in Fig.\ \ref{fig:str_vs_R}. One can see that indeed the trend in variation of $m$ and $l$ bonds is the same (the amplitude is larger) as in the compressively strained films (o-GdMnO$_3$ and o-TbMnO$_3$ films) with larger unit cell volumes. In this case, however, the inter-plane Mn-O-Mn bond angles are also reduced from their bulk values. 

\begin{figure}
\includegraphics[width=0.8\linewidth,trim=0.5cm 0cm 0cm 0cm]{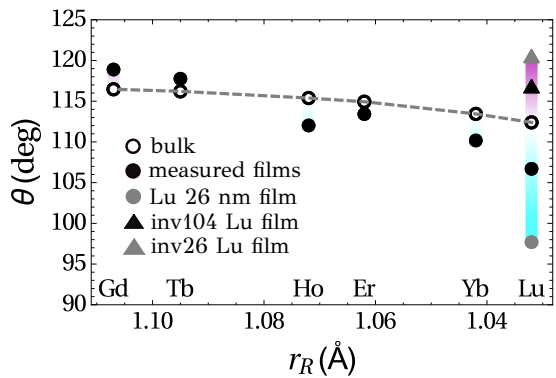} \\
\caption{\label{fig:theta_vs_R}Orbital mixing angles $\theta$ in strained films and bulk o-$R$MnO$_3$ versus the radius of the $R$ cation $r_R$. Bulk samples are shown by empty circles, experimentally measured strained films by filled circles. For o-LuMnO$_3$ the triangles denote calculations for hypothetical films which are compressively strained in the $ac$ plane by the same amount but in the opposite direction as the experimentally measured tensile strained films. The 26 nm o-LuMnO$_3$ film and the corresponding inverse case are highlighted in gray, the 104 nm film and the inverse case in black.  Compressive strain within the $ac$ planes of the o-$R$MnO$_3$ films is shown by the violet color, tensile strain by the blue color. The dashed lines connecting the data points for bulk o-$R$MnO$_3$ are guides to the eye.}
\end{figure}

To understand how these lattice variations affect the orbital ordering in o-$R$MnO$_3$, we estimate the orbital mixing angles $\theta$ using Eq.\ \ref{eq:orb_mixing} and our optimized values of $s$, $m$ and $l$ Mn-O bond lengths for all considered bulk samples and films of o-$R$MnO$_3$. The calculated $\theta$ are presented in Fig.\ \ref{fig:theta_vs_R}. One can clearly see that, since the Mn-O bond lengths are almost constant across the series of bulk samples, $\theta$ also shows only small variations. By applying strain, however, the orbital mixing angles can be significantly changed with respect to the corresponding bulk values. For example, for bulk LuMnO$_3$ $\theta$$\approx$112$^o$ and, according to Eq.\ \ref{eq:orb_mix_state}, the occupied $e_g$ orbitals on neighboring Mn sites $i$ and $j$ within the $ab$ planes have a character close to either $|3x^2-r^2\rangle$ (on site $i$) or $|3y^2-r^2\rangle$ (on site $j$). For the 26 nm film of LuMnO$_3$, however, $\theta$ is significantly reduced (97$^o$), which affects the character of the occupied orbitals, with the weight of $|3z^2-r^2\rangle$ state increasing and that of $|x^2-y^2\rangle$ going down, see Eq.\ \ref{eq:orb_mix_state}.      

Thus we see that chemical pressure and epitaxial strain are accommodated by the crystal structure of o-\textit{R}MnO$_3$ in different ways. In particular, the former leads to a change in the Mn-O-Mn bond angles (GFO distortion) while the latter affects mostly the Mn-O bond lengths (JT distortion) in the opposite way for compressive and tensile cases. Variation of the JT distortion in o-\textit{R}MnO$_3$ films changes their orbital ordering compared to bulk samples. Since the magnetism in the o-\textit{R}MnO$_3$ is closely related to the magnitudes of the JT and GFO distortions (as described in detail in Sec.\ \ref{sec:structure}),  the fact that the chemical pressure and epitaxial strain affect these distortions differently can be key to understanding of distinct magnetic (and, therefore, ferroelectric) properties of bulk and strained films of o-\textit{R}MnO$_3$. 

\section{Calculated effects of chemical pressure and epitaxial strain on the magnetism}
\subsection{Microscopic exchange interactions}
\label{sec:couplings}
In order to develop better insight into how these structural variations due to chemical pressure and epitaxial strain affect the magnetic properties of o-\textit{R}MnO$_3$, we analyze their effects on the microscopic exchange interactions. We extract all the considered Heisenberg, biquadratic and four-spin ring exchanges as well as the parameters of DMI and SIA (see Sec.\ \ref{sec:hamiltonian} and Fig.\ \ref{fig:exchanges}) by mapping the DFT energies of different magnetic configurations calculated for all the studied bulk o-\textit{R}MnO$_3$ (\textit{R}=Gd, Tb, Ho, Er, Yb, Lu) on the model Hamiltonian of Eq.\ \ref{eq:fullHam}. The methods which we use to extract the Heisenberg, biquadratic and four-spin ring exchanges are described in detail in our previous work (see Sec.\ IVB of Ref.\ \onlinecite{fedorova2015biquadratic}), while for calculations of DMI and SIA we employ the approach proposed in Sec II C of Ref.\ \onlinecite{xiang2011whangbo}. We show the extracted couplings $J_c$, $J_{ab}$, $J_b$, $J_{3nn}$, $B_{ab}$ and $K_c$ versus the radius of the $R$ cations in Fig.\ \ref{fig:JsvsR} (plots for the other coupling constants are presented in Fig.\ 1 of the Supplemental materials and the exact values of all the extracted couplings are summarized in Table IV of the Supplemental materials.) 

\begin{figure*}

\includegraphics[width=1\linewidth,trim=0cm 0cm 0cm 0cm]{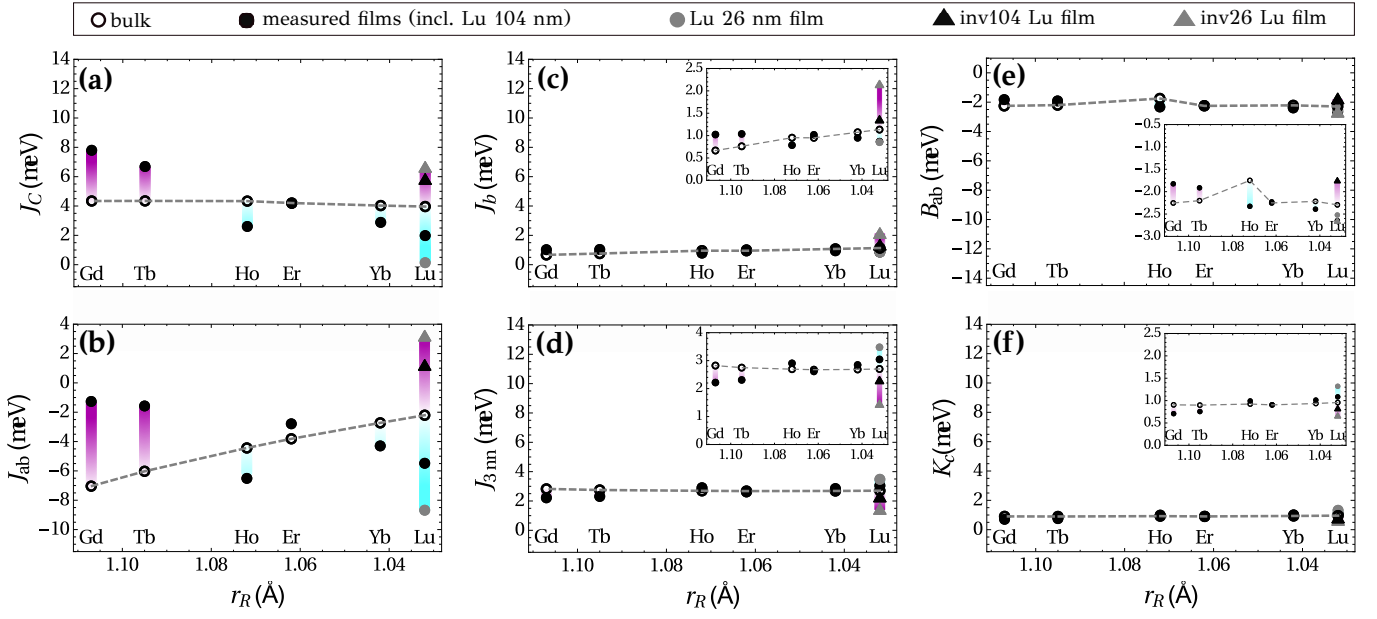}
\caption{\label{fig:JsvsR} Calculated exchange coupling constants versus the radius of the $R$ cation: (a)-(d) show the Heisenberg exchanges $J_c$, $J_{ab}$, $J_b$ and $J_{3nn}$, respectively; (e) shows the biquadratic in-plane exchanges $B_{ab}$ and  
(f) the four-spin ring couplings $K_c$. Bulk samples are indicated by the empty circles, experimentally measured strained films by the filled circles. For o-LuMnO$_3$ the triangles denote the hypothetical films which are compressively strained in the $ac$ plane by the same amount but opposite direction as experimentally measured tensile strained films. The o-LuMnO$_3$ 26 nm film and the corresponding inverse case are highlighted in gray, the 104 nm and inv104 nm cases in black. Compressive strain within the $ac$ planes of the o-$R$MnO$_3$ films is shown by the violet color, tensile strain by the blue color. The dashed line connecting the data points for the bulk o-$R$MnO$_3$ is used to guide the eye.}
\end{figure*}

One can see that the decreasing radius of the $R$ cation from Gd to Lu (resulting in the reduction of the Mn-O-Mn bond angles) in bulk o-$R$MnO$_3$, leads to a drastic decrease in the absolute value of the FM NN Heisenberg exchange $J_{ab}$ from -7.04 meV in o-GdMnO$_3$ to -2.20 meV in o-LuMnO$_3$ (Fig.\ \ref{fig:JsvsR} (b)). In contrast, all the other couplings remain almost constant across the series. The drop in $J_{ab}$ can be explained by the significant reduction of the FM contribution from the $e_g$-$p_\sigma$-$e_g$ superexchange, which is strongly dependent on the Mn-O-Mn bond angles. In contrast, the AFM $t_{2g}$-$p_\pi$-$t_{2g}$ contribution remains unchanged since it is much less affected by the variation of the bond angles due to the geometry of the participating orbitals. The latter also explains why the inter-plane NN Heisenberg couplings $J_c$ are nearly the same for all the considered bulk samples of o-\textit{R}MnO$_3$ as these couplings are mostly determined by the $t_{2g}$-$p_\pi$-$t_{2g}$ superexchange. Clearly, there is also a small effect on the NNN Heisenberg coupling $J_b$ (see the inset in Fig.\ \ref{fig:JsvsR} (c)), which increases with reducing \textit{R}. This occurs because of the decrease in the distance between the ions O(1) and O(2) shown in Fig.\ \ref{fig:rmno3_structure} (b), which results in larger overlap between their $p$ orbitals along the Mn-O(1)-O(2)-Mn superexchange path. We can conclude that the evolution of the magnetic order in bulk o-\textit{R}MnO$_3$ is mostly due to the reduction of $J_{ab}$, because the effect of other couplings (NNN Heisenberg, higher order couplings and anisotropic terms) becomes more pronounced when the strong FM NN exchange is reduced.

Next, we perform similar calculations of the exchange coupling and anisotropy constants for the films of o-\textit{R}MnO$_3$ to determine how they are influenced by the structural variations caused by epitaxial strain. The resulting couplings are presented in Fig.\ \ref{fig:JsvsR}; see also Fig.\ 1 and Table V of the Supplemental materials. As we showed in the previous section, the application of strain affects the Mn-O bond lengths, whereas the Mn-O-Mn bond angles in most cases change only slightly from their values in the corresponding bulk samples. First we consider four films which are compressively strained within the $ac$ plane (o-GdMnO$_3$, o-TbMnO$_3$, o-LuMnO$_3$ inv26 and inv104) and for which the $b$ lattice constants are expanded, resulting in a reduction of $m$ and increase in $l$ Mn-O bond lengths compared to the bulk samples. As one can see from Figs. \ref{fig:str_vs_R} (d) and \ref{fig:JsvsR} (a), the decrease in $m$ by 0.02 - 0.04 {\AA}  provides a significant increase in the coupling $J_c$ (for example, by 3.46 meV for a thin film of o-GdMnO$_3$ relative to the corresponding bulk sample). This can be explained by the increased overlap between the $d$ orbitals of Mn and $p$ states of O participating in the superexchange. The increase in $l$ (by 0.02-0.04 {\AA}, Fig.\ \ref{fig:str_vs_R} (e)), in turn, results in a drastic reduction in the absolute value of $J_{ab}$ coupling relative to the bulk samples for o-GdMnO$_3$ and o-TbMnO$_3$, and for o-LuMnO$_3$ films this coupling even changes sign from FM to AFM (see Fig.\ \ref{fig:JsvsR} (b)). The latter likely occurs because the AFM contribution from the $t_{2g}$ states start to dominate over the FM $e_g$ contribution. The increase in the NNN coupling $J_b$ (see Fig.\ \ref{fig:JsvsR} (c)) originates from the reduction of O(1)-O(2) distance, which is a secondary effect of the increase in $l$. Interestingly, the higher order couplings ($K_c$ and $B_{ab}$) are affected by the variation of the bond lengths while they show almost no dependence on the bond angles (see insets in Figs.\ \ref{fig:JsvsR} (e) and (f)).  
For the tensile strained films, the variation of the couplings is opposite to the case of compressive strain. For example, in the o-LuMnO$_3$ 26 nm film, $J_c$ is reduced to almost 0 meV due to the increase of $m$ Mn-O bond lengths and $J_{ab}$ is increased in absolute value to -8.7 meV, which is even stronger than the same coupling in bulk o-GdMnO$_3$, by decreasing $l$. The four-spin ring inter-plane exchange $K_c$ increases with tensile strain and starts to compete with the weak $J_c$. 

Thus we demonstrated that the microscopic exchange interactions in o-\textit{R}MnO$_3$ evolve differently under chemical pressure and epitaxial strain. Specifically, the substitution of smaller \textit{R} in bulk o-\textit{R}MnO$_3$ results in an increased GFO distortion and leads to the reduction of the NN in-plane Heisenberg coupling $J_{ab}$ and a slight increase in the NNN coupling $J_b$, while all the other couplings are almost constant across the series of the bulk samples. On the other hand, the change in the Mn-O bond lengths caused by epitaxial strain affects strongly both in-plane and inter-plane NN Heisenberg exchanges, and leads to a smaller variation of other coupling constants (NNN Heisenberg, biquadratic and four-spin ring exchanges).  
The changes are clearly different for compressive and tensile strain. The evolution of each coupling depends on whether the structure and, consequently, the Mn-O bond lengths are expanded or reduced in the relevant direction.

\subsection{Monte Carlo simulations}
\label{sec:montecarlo}
In the next step we perform a series of Monte Carlo simulations using the calculated exchange couplings and anisotropy constants  for bulk samples and strained films of o-\textit{R}MnO$_3$ to determine their ground state magnetic phases. This also serves as a check of how well the model Hamiltonian of Eq.\ \ref{eq:fullHam} reproduces the experimentally measured magnetism in these systems. 

\begin{figure}
\centering
\includegraphics[width=0.43\textwidth]{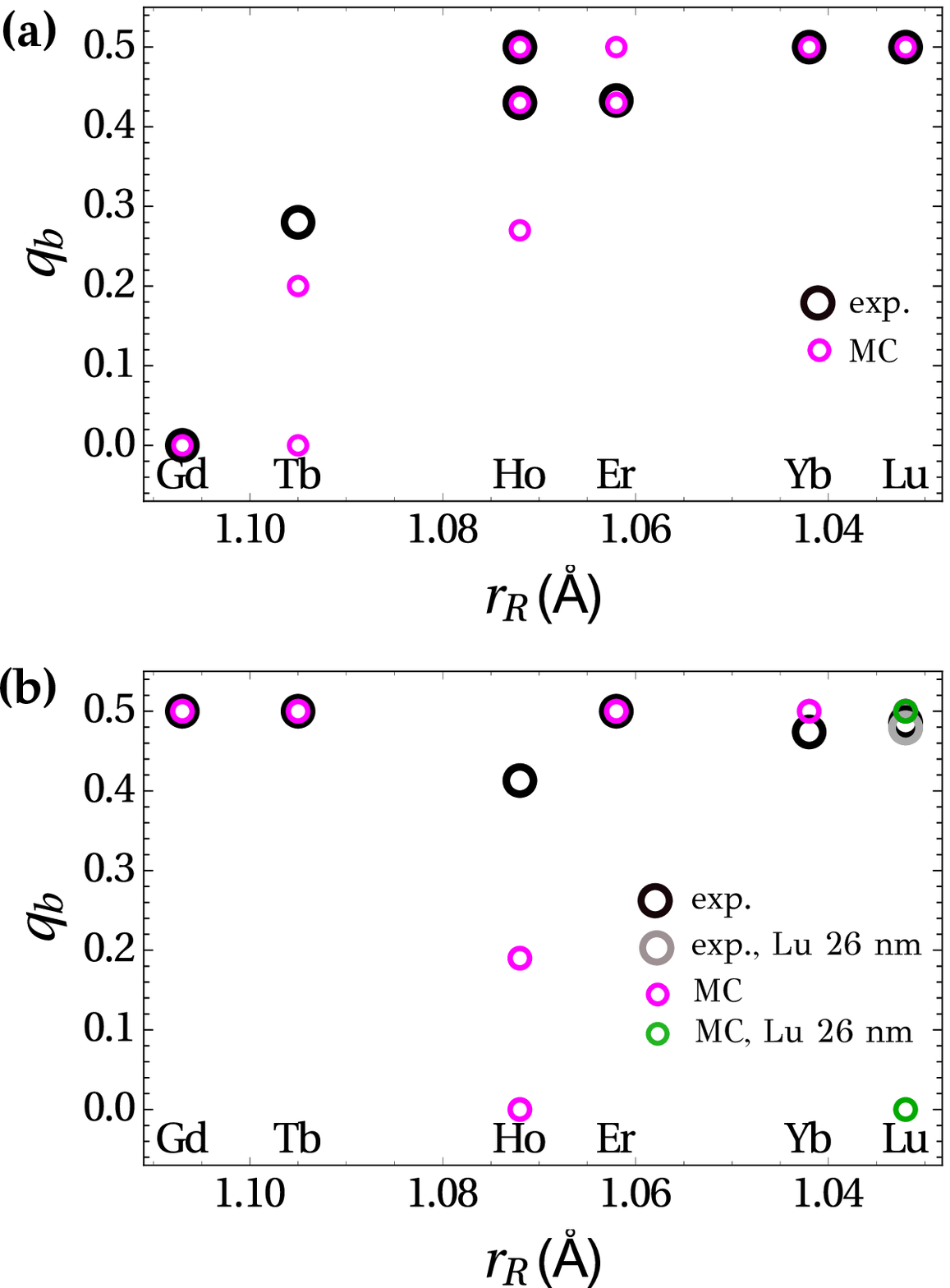}
\caption{\label{fig:mc_results} Experimentally determined and calculated modulation vectors of the ground state magnetic phases in bulk and strained o-\textit{R}MnO$_3$. (a) shows $q_b$ for bulk o-\textit{R}MnO$_3$, (b) for the films of o-\textit{R}MnO$_3$. Black circles indicate experimentally determined (exp.) $q_b$, purple circles calculated $q_b$ (MC). The gray circle in (b) indicates the experimentally measured $q_b$ in the 26 nm film of LuMnO$_3$, and the green circle shows the calculated $q_b$ for this film; the $q_b$ values for the 104 nm LuMnO$_3$ film are shown with the usual black (measured) or purple (calculated with MC) circles (note that the purple circle at $q_b$=0.5 is obscured by the green circle). $q_b$ is in reciprocal lattice units.}
\end{figure}

\begin{table}[t]
\caption{DFT energies per spin (in meV) (relative to the energy of the E-AFM order) calculated for bulk and strained o-LuMnO$_3$ (26 nm and 104 nm films) imposing E-AFM, H-AFM and I-AFM orders.}
\begin{tabular}{p{36pt}p{65pt}p{65pt}p{65pt}}
\hline
\hline
& \centering{E-AFM} & \centering{H-AFM} & \centering{I-AFM}
\tabularnewline
\hline
\centering{bulk} & \centering{0} & \centering{1.75}  & \centering{0.93}
\tabularnewline
\centering{104 nm} & \centering{0} & \centering{-0.68} & \centering{-1.78}
\tabularnewline
\centering{26 nm} & \centering{0} & \centering{-3.56} & \centering{-5.18}
\tabularnewline
\hline
\hline
\end{tabular}
\label{table:DFT_en}
\end{table} 

First, we consider bulk o-\textit{R}MnO$_3$ and determine the magnetic ground states for the systems with $R$=Gd, Tb, Ho, Er, Yb and Lu using the exchange coupling and anisotropy constants listed in Table  IV of the Supplemental materials. Since the methods which we use to calculate these constants allow an uncertainty in their values of up to $\pm$10-25\% (see our previous work for details, Ref.\ \onlinecite{fedorova2018fourspin}), we take the lower boundary of this uncertainty range and check whether the experimentally observed magnetic ground states can be reproduced for all systems within this range of parameters.
For that purpose we perform for each compound a set of MC simulations in which one of the exchange couplings ($J_c$, $J_{ab}$, $J_a$, $J_{diag}$, $J_b$, $J_{3nn}$, $K_{ab}$, $K_c$, $B_{ab}$, $B_c$, $\gamma_{ab}$, $\gamma_c$) or anisotropy ($A$) presented in Table IV of the Supplemental materials is varied by $\pm$10\% while all the others are kept fixed to the values presented in Table IV of the Supplemental materials. In these simulations the system size is 4$\times$100$\times$4 MC unit cells. For each compound, the lowest energy state obtained in the MC simulations with the couplings and anisotropy constants listed in Table IV of the Supplemental materials is used as a starting configuration. We determine the types of obtained magnetic phases by calculating the order parameters (for A-AFM, E-AFM and H-AFM orders) and magnetic structure factors along different directions in reciprocal space; the positions of the peaks in the magnetic structure factors give the modulation vectors of the resulting magnetic phases. 

In Fig.\ \ref{fig:mc_results} (a) we present the modulation vectors $q_b$ of the minimum energy phases obtained in our MC simulations for bulk o-\textit{R}MnO$_3$ together with the experimentally reported values.
We find that for o-\textit{R}MnO$_3$ with \textit{R}=Gd, Ho, Er, Yb, Lu our model Hamiltonian (Eq.\ \ref{eq:fullHam}) and the calculated couplings reproduce well the experimentally reported $q_b$ values. For o-TbMnO$_3$ we obtain a spiral order with $q_b=0.2$ as the lowest energy state (the experimental value is $q_b=0.28$) using periodic boundary conditions in all directions, while with open boundary conditions along the $b$ axis we obtain $q_b=0.22$. Interestingly, for $R$=Tb, Ho and Er several magnetic phases can be stabilized by varying the exchange couplings by $\pm$10\% of their values listed in Table IV of the Supplemental materials. This behavior is likely due to a competition between exchange interactions in these compounds (almost all calculated couplings are relatively strong), resulting in multiple low-energy magnetic states with very close energies. The favoring  of one state over another in the real samples may occur due to different synthesis conditions resulting in slightly different lattice parameters. For example, in o-TbMnO$_3$ samples, both A-AFM order and an incommensurate cycloidal spiral can be the lowest energy states. In o-HoMnO$_3$, in turn, a cycloidal spiral, w-spiral and E-AFM orders can be readily stabilized. The latter two can be the magnetic ground states in o-ErMnO$_3$ as well (see our previous work for the details, Ref.\ \onlinecite{fedorova2018fourspin}). This can explain the contradictory experimental reports of the magnetic and ferroelectric properties of the o-\textit{R}MnO$_3$ that are on the border between spiral and E-AFM phases in the magnetic phase diagram described in Sec.\ \ref{sec:phase_diagram}.

\begin{table}[b]
\caption{Electric polarizations (in $\mu$C/cm$^2$) calculated for bulk and strained GdMnO$_3$, ErMnO$_3$ and LuMnO$_3$ imposing E-AFM, H-AFM and I-AFM orders. The value of $P$ corresponding to the ground-state magnetic phase is in bold font.}
\begin{tabular}{p{36pt}p{65pt}p{65pt}p{65pt}}
\hline
\hline
& \centering{E-AFM} & \centering{H-AFM} & \centering{I-AFM}
\tabularnewline
\hline
& \multicolumn{3}{c}{GdMnO$_3$}
\tabularnewline
\hline
\centering{bulk} & \centering{4.17 $||a$} & \centering{0.08 $||c$}  & \centering{0.11 $||a$}
\tabularnewline
\centering{10 nm} & \centering{\textbf{3.16} $||a$} & \centering{0.31 $||c$} & \centering{0.17 $||a$}
\tabularnewline
\hline
& \multicolumn{3}{c}{ErMnO$_3$}
\tabularnewline
\hline
\centering{bulk} & \centering{4.06 $||a$} & \centering{0.35 $||c$} & \centering{0.12 $||a$}
\tabularnewline
\centering{30 nm} & \centering{\textbf{3.85} $||a$} & \centering{0.36 $||c$} & \centering{0.17 $||a$}
\tabularnewline
\hline
& \multicolumn{3}{c}{LuMnO$_3$}
\tabularnewline
\hline
\centering{26 nm} & \centering{5.17 $||a$} & \centering{0.18 $||c$} & \centering{\textbf{0.77} $||a$}
\tabularnewline
\centering{104 nm} & \centering{4.60 $||a$} & \centering{0.16 $||c$} & \centering{\textbf{0.45} $||a$}
\tabularnewline
\centering{bulk} & \centering{\textbf{4.09} $||a$} & \centering{0.40 $||c$} & \centering{0.19 $||a$}
\tabularnewline
\centering{inv104} & \centering{3.54 $||a$} & \centering{0.10 $||c$} & \centering{0.05 $||a$}
\tabularnewline
\centering{inv26} & \centering{3.22 $||a$} & \centering{0.06 $||c$} & \centering{0.26 $||a$}
\tabularnewline
\hline
\hline
\end{tabular}
\label{table:polarization}
\end{table} 

Next, we perform the same analysis for the strained films of o-\textit{R}MnO$_3$. The modulation vectors of the ground state magnetic phases obtained in our MC simulations and the corresponding experimental values are presented in Fig.\ \ref{fig:mc_results} (b). We find that for o-GdMnO$_3$, o-TbMnO$_3$ and o-ErMnO$_3$ films the lowest energy magnetic phase is E-AFM with the spins slightly canted away from the $b$ axis, which agrees with the experiments \cite{shimamoto2017phase_diagram,shimamoto2017tbmno3}. For the o-LuMnO$_3$ 104 nm and 26 nm films, the experimentally reported $q_b$ values are 0.486 and 0.479, respectively. In our MC simulations we observe magnetic phases with similar incommensurate $q_b$ values for these films, however we find these phases to be metastable. For the 104 nm film the calculated lowest energy state is H-AFM order with $q_b$=0.5. Note, that H-AFM is degenerate with I-AFM order with $\mathbf{q}$=(0,0.5,0.5) within the framework of the model Hamiltonian of Eq.\ \ref{eq:fullHam}, however the latter state does not give the experimentally observed peak in the magnetic structure factor at (0,$q_b$,0) (see Ref.\ \onlinecite{fedorova2018fourspin}). For the 26 nm film both H-AFM (or I-AFM) and A-AFM states can be stabilized in the simulations. The presence of the H-AFM (or I-AFM) order in these films of o-LuMnO$_3$ is interesting since one would rather expect the establishment of A-AFM order in o-$R$MnO$_3$ with such a strong NN in-plane Heisenberg exchange $J_{ab}$ (-5.48 and -8.68 meV, respectively). H-AFM (or I-AFM) order is enabled due to drastic suppression of the NN inter-plane Heisenberg coupling $J_c$ combined with an increased inter-plane four-spin ring interaction $K_c$ which favors this order. 
The only sample for which we did not reach an agreement with experiment is the o-HoMnO$_3$ film.  The experimental value of $q_b$=0.413 was not found even in the range of the couplings of $\pm$30\% of the values presented in Table V of the Supplemental materials. We believe that this is due to an experimental limitation. The low homogeneity of this sample likely causes inconsistencies between the RXD and XRD experiments, as they may probe slightly different positions and volumes of the sample.

To double check the results of our MC simulations for the films of o-LuMnO$_3$ (26 nm and 104 nm) in which unconventional H-AFM or I-AFM orders were obtained as the lowest energy states, and to clarify whether one of these states might be favored in these systems by, for example, exchange striction or another distortion of the electronic density, we perform the following analysis: We construct a 1$\times$2$\times$2 supercell for each film (the theoretically optimized unit cell is doubled along \textit{b} and \textit{c} directions) and optimize the ionic positions within this supercell imposing E-AFM, H-AFM and I-AFM orders in turn. Then we calculate the energies of these with their corresponding magnetic orders. The results are presented in Table \ref{table:DFT_en}. For comparison, the corresponding energies calculated for bulk o-LuMnO$_3$ are also presented. One can see that E-AFM order is the lowest energy state for bulk o-LuMnO$_3$. Tensile strain along the \textit{a} and \textit{c} directions 
favors the establishment of I-AFM order in both the 104 nm film and 26 nm films. 

Thus we showed that MC simulations based on the model Hamiltonian of Eq.\ \ref{eq:fullHam} and the exchange couplings calculated using DFT accurately reproduce the experimentally determined magnetic phase diagram of both bulk and strained o-\textit{R}MnO$_3$. We find that, in those bulk o-\textit{R}MnO$_3$ that lie near the boundary between IC spiral and E-AFM phases, different magnetic orders can be stabilized by small variations ($\pm$10\%) of the exchange interactions. In real materials, such small variations could arise from slightly different lattice constants due to different synthesis conditions and/or the presence of defects. This could explain the contradictory values reported for the measured magnetism and ferroelectricity in these materials. Our simulations also confirmed that E-AFM can be stabilized in o-GdMnO$_3$, o-TbMnO$_3$ and o-ErMnO$_3$ by epitaxial strain. Finally, we discovered an unconventional I-AFM order, which is degenerate with H-AFM order in the MC simulations, but lower in energy in DFT, in the 26 nm and 104 nm films of o-LuMnO$_3$. This order is enabled by the increased inter-plane four-spin ring exchange interactions $K_c$ and drastically reduced inter-plane NN Heisenberg couplings $J_c$ caused by the longer $m$ Mn-O bonds.    

\section{Electric polarization in bulk and strained \lowercase{o}-RM\lowercase{n}O$_3$}
\label{sec:polarization}

Finally, in order to understand how the chemical pressure and epitaxial strain affects the electric polarization in o-\textit{R}MnO$_3$, and to check whether the presence of the magnetic phases which were obtained in our MC simulations can resolve the contradictions in the reported measured values of $P$, we perform the following analysis: We consider bulk and strained films of o-GdMnO$_3$, o-ErMnO$_3$ and o-LuMnO$_3$ (for the latter both experimentally studied and hypothetical inv26 and inv104 films). For each system we construct 1$\times$2$\times$2 supercells by doubling the theoretically optimized unit cells along the \textit{b} and \textit{c} axes, and relax the ionic positions within these supercells imposing E-AFM, I-AFM and H-AFM orders. Then we perform Berry phase calculations using these relaxed structures with the corresponding magnetic orders imposed. The obtained polarizations, with the supercell in which the positions were relaxed with A-AFM order taken as the reference high-symmetry structure, are summarized in Table \ref{table:polarization}. One can see that $P$ calculated for bulk o-\textit{R}MnO$_3$ with E-AFM order imposed is almost unaffected by the size of the \textit{R} ion. All the values are at least an order of magnitude higher than those measured experimentally, in agreement with previous theoretical reports \cite{yamauchi2008rmno3,zhang2018rmno3}. (Note, that for \textit{R} larger than Gd, Refs. \onlinecite{yamauchi2008rmno3,zhang2018rmno3} reported an enhancement in \textit{P} with \textit{R}). Compressive strain along the \textit{a} and \textit{c} axes reduces \textit{P} in the films of E-AFM o-GdMnO$_3$ and inv104 and inv26 hypothetical films of o-LuMnO$_3$ by up to 1 $\mu$C/cm$^2$. Tensile strain along the same direction, in turn, increases $P$; for example, for the 26 nm film of o-LuMnO$_3$ $P$ increases by more than 1 $\mu$C/cm$^2$. Our calculated values are inconsistent with the recent experimental study of $P$ in the series of o-\textit{R}MnO$_3$ thin films, in which $P\approx1$ $\mu$C/cm$^2$ along the \textit{a} axis was reported for all \textit{R}=Gd,...,Lu except Tb, where $P\approx2$ $\mu$C/cm$^2$ was reported \cite{shimamoto2017phase_diagram}.

Our calculated $P$ values induced by H-AFM order are aligned along the $c$ axis and their amplitudes are at least an order of magnitude smaller than those induced by E-AFM order. While the absolute values of $P$ are less affected by structural modifications compared to the E-AFM case, the fractional changes are equally dependent. This direction of \textit{P}, has to our knowledge been experimentally observed only in systems with spiral magnetic orders (o-TbMnO$_3$ \cite{kimura2003magnetic,kimura2005magnetoelectric}, o-DyMnO$_3$ \cite{kimura2005magnetoelectric}, o-Eu$_{1-x}$Y$_{x}$MnO$_3$ \cite{hemberger2007multiferroic} and o-Gd$_{1-x}$Tb$_x$MnO$_3$ \cite{Yamasaki2008mixtures}), in bulk samples of o-HoMnO$_3$ \cite{lee2011mechanism} with incommensurate order ($q_b\approx$0.4) and in weakly strained films of o-YMnO$_3$ \cite{fina2010ymno3Pc}. In earlier work, Ref. \onlinecite{fedorova2018fourspin}, we showed that $P||c$ in o-HoMnO$_3$ can be explained by the presence of w-spiral order. 
I-AFM order induces a small polarization along the \textit{a} axis, with the value of $P$=0.77 $\mu$C/cm$^2$ that we obtain for the 26 nm o-LuMnO$_3$ film being close to the experimentally measured value of $P\approx$ 1 $\mu$C/cm$^2$ \cite{shimamoto2017phase_diagram}. The I-AFM phase, however, is the ground state only in the 26 nm and 104 nm o-LuMnO$_3$ films according to our MC and DFT calculations.

In conclusion, in spite of the fact that our DFT calculations correctly capture the various magnetic orderings in o-\textit{R}MnO$_3$ films and bulk samples, we are not able to reproduce the experimentally reported ferroelectric polarizations in many cases. The wide spread in the  reported values of ferroelectric polarizations in different samples of o-\textit{R}MnO$_3$, the consistently low values for E-AFM bulk crystals, as well as the similar values across the series of o-\textit{R}MnO$_3$ films remain unexplained.

\section{Summary and conclusions}
\label{sec:summary}

In summary, we studied the effects of chemical pressure and epitaxial strain on the crystal structure and multiferroic orders of the o-$R$MnO$_3$ series using X-ray diffraction measurement techniques (XRD and RXD), first-principles calculations and Monte Carlo simulations. 

In our RXD measurements we observed that the magnetic modulation vectors $q_b$ measured for o-\textit{R}MnO$_3$ films can differ significantly from those of bulk samples. To clarify the origin of this difference we used DFT to determine how the lattice parameters evolve in the o-$R$MnO$_3$ series, for both bulk and thin-film samples. We then studied the effect of these lattice variations on the microscopic exchange interactions. We found that reducing the radius of the $R$ cation in bulk o-$R$MnO$_3$ leads to decreasing Mn-O-Mn bond angles within the $ab$ planes and along the $c$ axis, while the Mn-O bond lengths stay almost constant throughout the series, with only the $l$ bonds decreasing slightly. In contrast, strain primarily affects the Mn-O bond lengths relative to the corresponding bulk samples, with bond angles varying under strain only in the samples with the smallest unit cell volumes. 

Next, we showed that reduction of the Mn-O-Mn bond angles due to decreasing $R$-cation radius in bulk o-$R$MnO$_3$ leads to a significant decrease in the absolute value of NN Heisenberg in-plane exchange $J_{ab}$ (see Fig.\ \ref{fig:exchanges}) and a small increase in the NNN Heisenberg coupling $J_b$. All other couplings and anisotropies remain almost constant with respect to $R$ radius. From this finding we concluded that the evolution of the magnetic order across the bulk series is dominated by the reduction in $J_{ab}$, which makes the effect of the other couplings, such as NNN Heisenberg couplings, biquadratic and four-spin ring exchanges, DMI and anisotropies, more pronounced. For films of o-\textit{R}MnO$_3$, we demonstrated that variation of the Mn-O bonds by applying strain can have a drastic effect on both in-plane and inter-plane NN Heisenberg couplings ($J_{ab}$ and $J_c$, respectively), and the magnitudes of the NNN Heisenberg couplings ($J_b$ and $J_{3nn}$) and of higher order exchanges (biquadratic and four-spin ring exchanges) can also be affected. Expansion and compression of the Mn-O bonds have opposite effects on the magnitudes of the exchange couplings.

In our Monte Carlo simulations we determined the magnetic ground states of the model Hamiltonian of Eq.\ \ref{eq:fullHam} for bulk and strained o-\textit{R}MnO$_3$ using the extracted exchange coupling and anisotropy constants, and found that the calculated modulation vectors agree well with the available experimental data. We showed that in those bulk o-\textit{R}MnO$_3$ on the boundary between IC spiral and E-AFM states in the magnetic phase diagram (Fig.\ \ref{fig:PD_2_options}), different magnetic orders can be stabilized by small variations of the exchange couplings. This can explain the contradictory experimental reports of their magnetic and ferroelectric properties. For compressively strained o-GdMnO$_3$ and o-TbMnO$_3$ films we confirmed the reported evolution of the magnetic order to the E-AFM phase. This occurs due to a drastic reduction of the NN in-plane Heisenberg coupling $J_{ab}$ caused by the increasing length of the $l$ Mn-O bonds. For tensile-strained films of o-LuMnO$_3$ we found that suppression of the inter-plane Heisenberg coupling $J_c$ and increase in the four-spin ring coupling $K_c$ can stabilize exotic magnetic orders such as H-AFM or I-AFM, with I-AFM having the lower DFT energy. 

Finally, we used DFT to analyze how the electric polarization would evolve in bulk and strain o-\textit{R}MnO$_3$ if it were induced by one of the magnetic phases which we obtained in our MC simulations. The values of $P$ calculated on imposing E-AFM order were significantly larger than the experimentally measured values for both bulk and films of o-\textit{R}MnO$_3$, and in the latter case is highly strain dependent, increasing with tensile strain along the $a$ and $c$ directions and vice versa. This behavior, however, has not been reported experimentally, where measured $P$ values are similar for both compressively and tensile strained films. We find that the $P$ values calculated with I-AFM order imposed are closest to those measured experimentally. However, in our MC and DFT calculations I-AFM is the lowest energy phase only in the tensile strained films of o-LuMnO$_3$. Therefore, our findings cannot fully resolve the puzzling behavior of $P$ in o-\textit{R}MnO$_3$.

\section{Acknowledgements}

Experiments were performed at the X11MA and X04SA beamlines at the Swiss Light Source, Paul Scherrer Institut, Villigen, Switzerland. We thank the X11MA and X04SA beamline staff for experimental support. The financial support of PSI and the Swiss National Science Foundation (SNSF) is gratefully acknowledged. Y.W.W., and M.R. acknowledge support by SNSF Projects No. 137657, and No. CRSII2\_147606, respectively. Funding was also received from the SNSF's National Centers of Competence in Research, Molecular Ultrafast Science and Technology (NCCR MUST) and Materials’ Revolution: Computational Design and Discovery of Novel Materials (NCCR MARVEL). E.M.B. acknowledges funding from the European Community’s Seventh Framework Programme (FP7/2007-2013) under grant agreement No. 290605 (PSI-FELLOW/COFUND). A.A. acknowledges funding from the University of Fribourg. Financial support and CROSS funding to Yi Hu and Kenta Shimamoto from PSI are also acknowledged. 
N.S.F. and N.A.S. acknowledge the ERC Advanced Grant program (No.\ 291151) and ETH Z\"{u}rich for financial support. Computational resources were provided by ETH Z\"{u}rich and Swiss National Supercomputing Centre (CSCS), project No. p504.
We thank Andrea Scaramucci for providing the Monte Carlo code and for fruitful discussions.
  
\bibliographystyle{apsrev}
\bibliography{main.bib}

\end{document}